\documentclass[12pt,preprint]{aastex}
\usepackage{epsf}
\usepackage{lscape}

\usepackage{graphicx}
\usepackage{natbib}
\begin{document}

\title {SHELS: TESTING WEAK LENSING MAPS WITH REDSHIFT SURVEYS}

\author {Margaret J. Geller} 
\affil{Smithsonian Astrophysical Observatory,
\\ 60 Garden St., Cambridge, MA 02138}
\email{mgeller@cfa.harvard.edu}
\author {Michael J. Kurtz} 
\affil{Smithsonian Astrophysical Observatory,
\\ 60 Garden St., Cambridge, MA 02138}
\email{mkurtz@cfa.harvard.edu}
\author {Ian P. Dell'Antonio} 
\affil{Department of Physics, Brown University, 
\\Box 1843, Providence, RI 02912}
\email{ian@het.brown.edu}
\author{Massimo Ramella}
\affil{INAF, Osservatorio Astronomico di Trieste,
\\ via C. B. Tiepolo 11, I-34131 Trieste, Italy}
\email {ramella@ts.astro.it}
\author {Daniel G. Fabricant} 
\affil{Smithsonian Astrophysical Observatory,
\\ 60 Garden St., Cambridge, MA 02138}
\email{dfabricant@cfa.harvard.edu}

\begin {abstract}
Weak lensing surveys are emerging as an important tool for the construction of ``mass selected'' clusters of galaxies. We evaluate both the efficiency and completeness of a weak lensing selection by combining a dense, complete redshift survey, the Smithsonian Hectospec Lensing Survey (SHELS), with a weak lensing map
from the Deep Lens Survey (DLS). SHELS includes 11,692 redshifts for galaxies with R$\leq 20.6$ in the four square degree DLS field; the survey is a solid basis for identifying massive clusters of galaxies with redshift $z \lesssim 0.55$. The range of sensitivity of the redshift survey is similar to the range for the DLS convergence map. 
Only four  the twelve convergence peaks with signal-to-noise
$\geq 3.5$ correspond to clusters of galaxies with $M \gtrsim 1.7 \times 10^{14}$M$_\odot$. Four of the eight massive clusters in SHELS are detected in the weak lensing map
yielding a completeness of $\sim 50$\%. We examine the seven known extended cluster x-ray sources in the DLS field: three can be detected in the weak lensing map, three should not be detected without boosting from superposed large-scale structure, and one is mysteriously undetected even though its optical properties suggest that it should produce a detectable lensing signal. Taken together, these results underscore the need for more extensive comparisons among different methods of massive cluster identification.
\end{abstract}

\keywords {galaxies: clusters: individual (CXOU J091551+293637, CXOU J091554+293316,
CXOU J091601+292750, XMMU J091935+303155, A781, CXOU J09202+302938, CXOU J092053+302800, CXOU J092110+302751 ) --- galaxies: distances and redshifts ---
gravitational lensing --- large-scale structure of universe }

\maketitle

\section{Introduction}
\label{intro}

Weak lensing maps and redshift surveys are fundamental, complementary tools
of modern cosmology. A weak lensing map provides a  weighted ``picture''
of projected surface mass density; a redshift survey provides a
three-dimensional map of the galaxy distribution resolving structures along
the line of sight within the broad window ``imaged'' by the lensing map. 

Although galaxies are biased tracers of the mass distribution, comparison of a
weak lensing map with a foreground redshift survey covering the appropriate
range promises progress in resolving some of the issues limiting 
cosmological applications of weak lensing. Here we focus on the use of
weak lensing for identifying clusters of galaxies. 

We compare the set of massive clusters identified in a dense, complete redshift foreground redshift survey
with significant weak lensing map convergence peaks. The Smithsonian Hectospec Lensing Survey (SHELS hereafter : Geller
et al. 2005) includes redshifts for
11, 692 galaxies in a four square degree field of the Deep Lens Survey (Wittman et al. 2002; DLS hereafter). SHELS enables identification of massive clusters for redshifts $z \lesssim 0.55$, the range of maximum sensitivity of the
DLS. Comparison of the two surveys measures {\it both} the efficiency and completeness of the weak lensing map for the identification of massive clusters.  These combined measures are the first direct
assays for any weak lensing survey. They provide an important benchmark for future work. 

\subsection {History and Challenges}
\label{history}
The construction of large catalogs of clusters of galaxies has a long history in
the optical (e.g. Abell 1958; Zwicky et al. 1968; Abell et al. 1989; Koester et al. 2007) and in the x-ray (e.g. Schwarz 1978;
Piccinotti 1982; Edge et al. 1990; Ebeling et al. 1996; B\"ohringer et al. 2004). In 
these  approaches, the 
catalog selection parameters are related to the cluster
mass only through scaling relations. In contrast, weak lensing offers the enticing possibility
of obtaining a ``(projected) mass selected'' catalog of clusters directly. 

Several studies show that weak lensing maps are a route
to cluster identification. Wittman et al. (2001) used the Deep Lens Survey to
make the first detection of a previously uncatalogued cluster from a
convergence map. Subsequently Miyazaki et al. (2002, 2007), Hetterscheidt
et al. (2005), Wittman et al. (2006), Schirmer et al. (2007),
and Gavazzi \& Soucail (2007), Hamana et al. (2008), Berg\'e et al. (2008), and Dietrich et al. (2008) have all demonstrated coincidence of optical and/or x-ray clusters with peaks in weak lensing convergence maps.     

The number of candidate clusters identified from weak lensing maps has risen steeply
to several hundred. For the two largest sets of candidates,
the success rate for identifying these candidates with clusters
of galaxies differs substantially. Miyazaki et al. (2007) claim 
an 80\% success rate for the $\sim$ 100 peaks in their convergence maps
with a signal-to-noise greater than 3.69. Hamana et al. (2008) support this claim with sparse  spectroscopic sampling of  36 weak lensing cluster candidates. In contrast, Schirmer et al. (2007) 
identified 158 possible mass concentrations in a blindly selected sample
and derive a $\sim$45\% success rate for their $\sim$4$\sigma$ convergence 
map peaks consistent with an earlier evaluation of a subsample of the survey
(Maturi et al. 2007). The 4$\sigma$ significance threshold may be an overestimate because Schirmer et al (2007) combine sets of peaks identified with two different statistics to construct their sample.   A third independent analysis of  convergence maps
from the CFHTLS (Gavazzi \& Soucail 2007) yields an intermediate
$\sim$65\% success rate among 14 peaks identified 
above a signal-to-noise threshold of 3.5.

Miyazaki et al. (2007) attribute the much lower success rate of Schirmer et al. (2007)
to the generally lower source density in the Schirmer et al. (2007) weak lensing maps.
However, there are other  possibly important differences in the construction 
and analyses of the maps. Schirmer et al. (2007) include a detailed discussion of
a variety of statistical issues in both the identification of convergence map 
peaks and of the related concentrations of galaxies. Although there are some
qualitative aspects in their evaluation of the galaxy counts, they limit the
identification of coincidences to apparent systems with redshifts 
less than $\sim$ 0.3.
This limit is roughly consistent with their lower source density. In
contrast,
Miyazaki et al. (2007) and Hamana et al. (2009) claim probable coincidences to a redshift of 0.5 or more.
Short of performing exactly the same analysis on the two datasets it is 
difficult to account for the differing success rates from consideration of
the relative background source counts alone.

Miyazaki et al. (2007) use a combination of x-ray observations,
redshift measurements (extended by Hamana et al. (2009)), and imaging to identify their convergence peaks with
systems of galaxies. Schirmer et al. (2007) take a more uniform approach of 
counting galaxies in beams with a 2$^\prime$ radius around the centers of 
their convergence peaks; they measure the apparent overdensity in
the magnitude range R $\sim$ 17 -- 22 and make an assessment of significance
depending on the excess count (with some occasional qualitative modifications).

Gavazzi \& Soucail (2007) also
take a uniform approach to the evaluation of significance of the peaks; they
use the distribution of photometric redshifts for galaxies in a circular
aperture of 2$^\prime$ centered on each convergence peak. They subtract a 
background distribution computed from galaxies which lie more 
than 6$^\prime$ from any convergence peak. For 9 of their 14 peaks, they
define the associated cluster redshift as the location of the most prominent
photo{\it z} peak. One remarkable feature of the Gavazzi \& Soucail
analysis is the low velocity dispersion (as low as 450 km/s) for some of the   
systems corresponding to convergence peaks.

Most recently Kubo et al. (2009) use x-ray observations, DLS images, and
available redshifts to estimate the fraction of false positives in 
a maximum likelihood weak lensing reconstruction.  They conclude
that only 10-25\% of their peaks with signal-to-noise greater than 3.5
are false positives. This conclusion may be overly optimistic because they
consider coincidences of peaks and apparent systems with angular separations
as large as 4 arcminutes. Schirmer et al. (2007) and Gavazzi \& Soucail (2007)
show that the average 
angular offset between convergence peaks and optical counterparts is 
0.9$\pm$0.5$^{\prime}$. The convergence maps of Schirmer et al. (2007) and Gavazzi \& Soucail (2007) are sensitive to lensing systems
in approximately the same redshift range as the DLS.

Recent simulations of the efficacy of convergence maps for construction of 
mass-selected cluster catalogs are based on ray-tracing through large n-body
simulations (Hamana et al. 2004; Hennawi \& Spergel 2005). Hamana et al. (2004) make a set of mock observations where they 
associate convergence map peaks with halos. They conclude that for convergence
peaks with a signal-to-noise $>$ 4, more than 60\% of massive 
halos are detected.
On the other hand, they also estimate that a conservative 40\% of 
these convergence peaks are false positives. Although the details of the 
simulation differ from the observations, the estimated false positive rate is
consistent with the conclusions of Gavazzi \& Soucail (2007), more optimistic
than Schirmer et al. (2007), and less optimistic than Miyazaki et al. (2007)
or Kubo et al. (2009). The $\sim$80\% rate of the Miyazaki et al (2007) survey is very close to the $\sim$85\% maximum achievable (Hennawi \& Spergel). Simulations by Dietrich et al. (2008) also show that 75\%
of matches  between convergence map peaks and massive halos are within 2.15$^{\prime}$, consistent
with the observational offsets observed by Schirmer et al. (2007) and Gavazzi \& Soucail (2007). 

In all of these recent studies some convergence peaks are associated with 
superpositions of several systems, often with low velocity dispersion,
along the line-of-sight toward the weak lensing peak. A clean evaluation 
of the nature and frequency of such superpositions is
very difficult without a complete foreground redshift survey. The fraction of
``dark'' peaks attributable to noise appears to vary with the nature 
of the convergence map and with the method of 
cross-identifying convergence peaks with physical systems. Yet another issue
not considered in analyses to date is the frequency of accidental
coincidences between convergence map peaks and apparent systems of galaxies.
Understanding the noise properties of convergence maps and the related
correspondence between the weak lensing signature and structures
in the galaxy distribution is a continuing challenge (Mellier 1999; Schneider 2006).

\subsection {The Importance of a Foreground Redshift Survey}
\label{zintro}
A foreground redshift survey is an independent measure of the matter 
distribution revealed by a convergence map. 
Here we use a complete redshift survey, SHELS,  to compare
peaks in the DLS convergence map with probable systems identified in the
redshift survey. We develop procedures similar to those adopted by
Schirmer et al. (2007) and Gavazzi \& Soucail (2007) who use narrow probes 
through their photometric data to evaluate coincidences between potential lensing systems and peaks in their
convergence maps.

The advantages of a redshift survey for evaluating the efficiency of a lensing
include (1) resolution of structures along the line-of-sight,
(2) availability of a velocity dispersion, and (3) bases for constructing 
a control sample and evaluating false positives. Combining a redshift survey
with a lensing map enables an evaluation of the frequency of 
chance coincidences between significant lensing peaks and systems of galaxies or
halos. The redshift survey also provides an estimate of the number of systems
undetected by the weak lensing map. Consideration of these issues has 
previously been limited largely to simulations.

In Section 2 we describe the DLS  lensing map and the SHELS redshift survey. We calculate the sensitivity
of the DLS to clusters with a given rest frame line-of-sight velocity dispersion in Section \ref{sensitivity}. Section \ref{probes} develops the technique we use for evaluating the match between the
convergence map peaks and structures in the redshift survey.  We examine the redshift distribution along the line-of-sight toward the 12 most significant weak lensing peaks in 
Section \ref{shelscandidates} and compute rest  frame line-of sight velocity dispersions for these candidate DLS/SHELS clusters in Section \ref{vdisps}. We  discuss the seven known extended clusters x-ray sources in Section \ref{xraysF2} and construct a set of SHELS candidate clusters independent of the DLS map in Section \ref{shelsclusters}. We evaluate the efficiency and completeness of the
DLS convergence map in Section \ref{weakcluster}. In Section \ref{photoz} we demonstrate the potential impact of photometric redshifts on the evaluation of the efficiency of weak lensing for massive  cluster identification. Section \ref{discussion} compares our results with simulations by Dietrich et al. (2008) and Hamana et al. (2004).
We conclude 
in Section 7. We use the WMAP concordance cosmology (Spergel et al. 2007) 
throughout with H$_o$ = 73, $\Omega_m = 0.3$ and 
$\Omega_\Lambda =  0.7$.

\section{The Data}
\label{data}
We use two ambitious surveys to explore the coincidence  of lensing peaks with
halos observed as systems of galaxies with rest-frame line-of-sight velocity 
dispersions $\gtrsim 600$ km s$^{-1}$ or, equivalently, masses $\gtrsim 2\times10^{14}$\ M$_\odot$.
The DLS (Wittman et al. 2002) is an NOAO key program covering 20 square degrees
in five separate fields; we use the four square degree F2 field at 
$\alpha$ = 09$^h$19$^m$32.4$^s$ and 
$\delta$ = +30$^{\circ}$00$^{\prime}$00$^{\prime\prime}$. SHELS is a
redshift survey covering the F2 field. SHELS is 98\% complete to a limiting apparent
magnitude R = 20.3 and differentially 60\% complete in 
the interval R = 20.3 - 20.6 (Geller et al. 2005; Kurtz et al. 2010)

\subsection {The DLS Map (F2)}
\label{dlsdata}
Photometric observations of F2 were made with the 
MOSAIC I imager (Muller et al. 1998) on the KPNO Mayall 4m telescope between 
November 1999 and November 2004. The R band exposures, all taken in seeing
$ < 0.9^{\prime\prime}$ FWHM, are the basis for the weak lensing map of F2. The
total exposure is 18000 seconds; the 1$\sigma$ surface brightness limit in R is 28.7 mag arcsec$^{-2}$
yielding about 45 resolved sources per square arcminute.
Wittman et al. (2006) describe the reduction pipeline.

Kubo et al. (2009) carry out a maximum likelihood lensing reconstruction of F2.
They base the convergence ($\kappa$) map on sources in the range 22.0 $ <$ R $ < $ 25.5. After
removal of objects which are too small relative to the PSF, so large that
they are probably nearby, or so elongated that they are probably superpositions
(Wittman et al. 2006), the source catalog contains 328,000 galaxies or
23 galaxies arcmin$^{-2}$. This source density is within the range of
other ground-based surveys (Miyazaki et al. 2007; Schirmer et al. 2007; Gavazzi \& Soucail 2007). We use this map for our evaluation of the efficacy of
the Kubo et al. (2009) weak lensing maps for locating massive clusters.

Kubo et al. (2009) construct noise maps of F2 with 100 realizations based on the 
source positions in the original catalog. In each realization, we assign 
ellipticities randomly to sources. We define $\sigma_{DLS}$ as the rms deviation in the noise maps.
We define the signal-to-noise as $\nu = \kappa/\sigma_{DLS}$. At each point in the image, we produce
a $\nu$ map   by dividing the reconstructed $\kappa$ value by the rms variation in the random realizations at the same spatial positions. Kubo et al. (2009) select candidate shear peaks as local maxima in this 
$\nu$ map which has a pixel scale of 1.5 arcminutes/pixel.

In the F2 field of the DLS, Kubo et al. (2009) identify 12 peaks with 
 $\nu \gtrsim$ 3.5. Although Kubo et al. (2009) construct their list of potential peaks using SExtractor with a 9-pixel area filter, the peak significance for selecting the $\nu > 3.5$ peaks is determined by the signal-to-noise in the highest significance pixel. Kubo et al. (2009) also estimate a total signal-to-noise for their most significant peaks. However, their method cannot resolve overlapping peaks like those
corresponding to the two massive clusters comprising Abell 781 (see Geller et al. 2005). Thus  we do not use the Kubo et al. (2009) total signal-to-noise values  here.

Kubo et al. (2009) tentatively identify 10 of these weak lensing peaks with
plausible systems of galaxies. These cross-identifications are based on
inhomogeneous data in an approach similar to Miyazaki et al. (2007) and 
the success rate is similar. 
One concern in some of these identifications (also noted by Kubo et al. (2009)) 
is that the separation between the 
center of the convergence peak and the galaxy system is as large as 
4$^{\prime}$. 
Schirmer et al. (2007) and
Gavazii \& Soucail (2007) find that the typical separation of 
convergence peaks and plausibly associated galaxy systems is about an
arcminute.

\subsection{Lensing Sensitivity}
\label{sensitivity}
To estimate the completeness of the weak-lensing selected cluster 
catalog derived from the $\kappa$ map (Kubo et al. 2009), we calculate the sensitivity limits as
a function of redshift. For any individual background galaxy in the weak lensing limit, the 
induced {\it measured} tangential ellipticity (and hence the S/
N for detection) depends on both the angular diameter distance to the 
galaxy and on the size of the galaxy relative to the PSF.
For a collection of galaxies, the effective total weight as a function of redshift is a 
polarizability-weighted sum over the individual distance ratios.  The effective distance ratio for the ensemble of background galaxies is 

\begin{equation}
 W_{eff}(z_l) = {{\sum_{s}  \frac{D_A(0,z_l) D_A(z_l,z_s)}{D_A(0,z_s)}} W_s \over {\sum_{s} W_s }}
\end{equation}
\\
\noindent where the subscripts $s$ and $l$ refer to the sources and lens, respectively. $W_s$ is the weight of each source galaxy (which depends on the object size), $z_l$ is the lens redshift,  $D_A(z_1,z_2)$ is the angular diameter distance between $z_1$ and $z_2$, (the ratio is zero if $z_s< z_l$), and the 
sum is over all sources.  For a realistic survey  an additional suppression results 
from misidentification of faint foreground galaxies, each of which should 
have weight zero.  Furthermore, in a ground-based survey, a large 
number of faint galaxies are unresolved.  Kubo et al. (2009)
reject objects with FWHM smaller 
than 1.2 times the PSF size; thus $W_s=0$ for these objects.

For each redshift, we compute the angular diameter 
distance factors in equation (1)  
for each redshift assuming the ``concordance" 
cosmology (Spergel et al. 2007).  
We derive the $W_s$ terms  using data from the COSMOS Subaru galaxy 
catalog (Taniguchi et al. 2007) and the COSMOS ACS catalog (Leauthaud 
et al. 2007).  The COSMOS combined ACS/Subaru dataset is currently 
the best dataset for this calibration because it covers by far the 
largest area with both ground- and space-based resolution to a depth comparable 
to the DLS.  In addition, the Subaru multi-color observations provide 
photometric redshift estimates  (Mobasher et al. 2007) 
for the majority of galaxies in the survey area. We use the COSMOS catalog size and photometric redshift information to estimate the redshift distribution of the galaxies that would be resolved in DLS. The comparison of ground-based and space-based COSMOS images also provide a measure of how well resolved DLS galaxies are as a function of their DLS size. Thus the COSMOS data provide a route to
a proper weighting of the DLS galaxies as a function of their probable intrinsic size in equation (1).  We match galaxies  
in the COSMOS Subaru and ACS catalogs with the additional requirement that the catalog 
magnitudes in I  match to within 0.3 magnitudes (This restriction eliminates 
errors in matching the two catalogs; it also eliminates most objects that cannot be adequately separated from neighboring objects in the Subaru imaging).

To construct a galaxy sample equivalent to the DLS sample, we select
all galaxies from the COSMOS galaxy catalog with  Subaru r  
magnitude in  20 $<$ r $<$ 25.3 (a rough match to the DLS R magnitude), 
scale radius $> 0.3^{\prime\prime}$  (to ensure 
that the measured size when convolved 
with the DLS PSF is larger than the size cutoff in the DLS catalogs), 
and $0.01<z_{phot}<2.5$.  The last cut  removes some  potentially catastrophic 
redshift errors; we choose the high-redshift cutoff  $\gg z_{max}$, 
the maximum lens redshift considered, to avoid biasing the high-redshift 
tail of the sensitivity. The cut on size produces an effective  mean source redshift in the
range 0.7 --- 0.8. MMT spectroscopy of a small subset of the DLS sources is consistent with this estimate (Geller et al. 2005). Approximately 29,000 COSMOS objects meet all 
of the criteria yielding a source density comparable (but 
about 15\% lower) with the source density of objects  in the weak 
lensing reconstruction of DLS field F2.   For each object, we compute 
the FWHM 
convolved with the mean PSF for the DLS data, 
and we compute the weight for that size (and PSF) as in Wittman et al. (2006).  For each lens redshift, we compute $W_{eff}$.

Because we identify clusters from the weak lensing map based on the projected surface density within a single reconstruction pixel (1.5$^\prime$x1.5$^\prime$), the S/N should scale linearly with the mass enclosed within that surface area.  Thus, we normalize the redshift dependence of our sensitivity by comparing to simulations. To calculate the absolute sensitivity of the weak-lensing survey to massive halos, we follow a modified version of the procedure for generating simulated catalogs in Khiabanian \& Dell'Antonio (2008).  We generate simulated galaxies with the same size-magnitude and magnitude-redshift relations as the HDF North and South fields (Williams et al. 1996; Casertano et al. 2000) using the prescription of
Khiabanian \& Dell'Antonio (2006). We use these simulated galaxies to
populate seven logarithmically-spaced redshift shells. We then distort the images with a lens (modeled as a core-softened cutoff isothermal sphere with a given rest frame line-of-sight velocity dispersion, $\sigma_{iso}$, a 1$^{\prime\prime}$ core radius, and a 100$^{\prime\prime}$ cutoff radius.  We resample the distorted images with the MOSAIC pixel scale, convolved with the model DLS PSF, and add noise to match the noise in the F2 image.  We repeat this procedure  for different values of $\sigma_{iso}$ (at a fixed $z_{lens}=0.3$) to generate maps of the weak lensing signal.  We run the Kubo et al. (2009) detection method  on these images to determine the value of $\sigma_{iso}$ where the peak S/N reaches the $\nu = 3.5$ cluster selection limit in Kubo et al. (2009).  In Figure \ref{fig:sensitivity_band.ps}, we plot the detection limit as a function of redshift ($L(z) = L(z
=0.3) {W_{eff}(z=0.3)/W_{eff}(z)}$); the solid line represents $\nu = 3.5$ and the dotted lines represent the detection limits for $\nu = 3$ and
$\nu = 4$, respectively. 
Because the detection criteria for the simulated clusters match the Kubo et al. (2009) sample, the simulation may overpredict the signal from clusters at very low redshift ($z<0.1$);  the angular region  where we calculate the potential in the reconstruction encloses significantly less mass (or alternatively, we spread the same signal over multiple pixels).

There are some fundamental limitations on the accuracy
of our survey sensitivity
calculations.  First, and most important, 
there is cosmic variance in the number of background galaxies 
behind a cluster.  For clusters near the detection limit, the shear 
signal is detectable in a region containing only 1000-2000 galaxies; 
on those scales there are  myriad effects that can greatly alter 
the number of background galaxies.  Second, although the COSMOS/Subaru 
field is extremely valuable, it covers $\lesssim 1$ square degree of the 
sky.  Therefore, any systematic differences between the properties 
of the COSMOS field and DLS F2 can change the
normalization.  One of the results of the analysis of the different 
DLS fields is that there are strong field-to-field variations in the 
types and amounts of structure present on degree scales. Without large 
surveys, there is no way around this fundamental limitation.

There are also several more accessible sources of uncertainty 
in the sensitivity calibration technique. 
First, the computed mass sensitivity  depends somewhat on the  model 
mass profile;
more centrally concentrated clusters are more easily detected.  Second, 
because we select galaxies in the COSMOS/Subaru r band, 
which is not identical to the DLS R band, there is a slight bias in 
the redshift and size distributions of the COSMOS galaxies.  Finally, 
the photometric redshift estimates introduce a bias if the faint 
galaxies have a significantly different redshift distribution from 
the brighter ones.  However, all these sources of uncertainty 
change the calibration by less than $\sim 10$ \%.  For example, 
taking the magnitude cutoff in the COSMOS catalog as 24.8 instead of 25.3, 
a much larger change than the uncertainty in the magnitude calibration, 
changes the redshift of peak sensitivity by 0.01, and changes 
the sensitivity at z=0.4 by 4\%.  In the analysis of the incompleteness 
of weak lensing cluster detection, we take a 10\% uncertainty in the
sensitivity limit into account when determining which clusters 
selected from the redshift survey should be detected in the lensing map.

\subsection {SHELS}
\label{shelsdata}

We investigate 
the association between convergence peaks and clusters (halos) in the galaxy 
distribution based on a complete redshift survey.  Our goal is identification of the systems of galaxies 
that should produce a weak lensing signal based on Figure \ref{fig:sensitivity_band.ps}. We thus 
identify systems in the redshift survey with rest frame 
line-of-sight velocity dispersion $\gtrsim 500$ km s$^{-1}$ (see Section \ref{shelsprobes}).

We constructed the galaxy catalog from the R-band source list for the F2 field of the
DLS.
We used surface brightness to separate stars from galaxies (see Kurtz 
et al. 2010 for details). The final catalog
contains 9825 galaxies with R $\leq$ 20.3; 9603 of these galaxies have redshifts.

We acquired spectra for the objects with the Hectospec 
(Fabricant et al. 1998, 2005) on the MMT from
April 13, 2004 to April 20, 2007. The Hectospec observation
planning software (Roll et al. 1998) 
enables  efficient acquisition of a magnitude limited sample.

The SHELS spectra cover the wavelength range 3500 --- 10,000 \AA
\ with a resolution of $\sim$6 \AA. Exposure times ranged from 0.75 --- 2 hours.
The lowest surface brightness objects in the survey required the longer 
integrations. We reduced the data with the standard Hectospec pipeline
(Mink et al. 2007) and derived redshifts with RVSAO (Kurtz \& Mink 1998) with
templates constructed for this purpose (Fabricant et al. 2005). Our 1151
galaxies with repeat 
observations yield robust estimates of the median error  in $cz$ where $z$ is the redshift. For emission line objects, the median error (normalized by $(1 + z)$) is 27 km s$^{-1}$; the median for absorption line objects (again normalized by $(1 + z)$) is 
37 km s$^{-1}$.
(see also Fabricant et al.  2005).

The integral completeness of SHELS to R = 20.3 is 97.8\%; the differential completeness
at the limiting magnitude is 94.6\%. The 218 objects (out of 9825) without redshifts
are low surface brightness blue objects or objects near the survey 
corners and edges.
Figure \ref{fig:completeness} shows 
the completeness of SHELS as a function of apparent magnitude
(note that the completeness scale ranges from 0.9 to 1.0).

The SHELS survey also includes 1871 galaxies with $20.3 < R \leq 20.6$;
the survey is $\sim 60$\% complete in this magnitude interval. The completeness is
patchy across the field but is generally at or above the mean in the region
of the most significant weak lensing peaks.

The median redshift of SHELS is 0.295 for the magnitude limited sample with R $\leq$ 20.3. Figure \ref{fig:zhisto} shows 
the redshift distribution
for the survey to a limiting R = 20.3. The impact of large-scale structure is obvious to 
a redshift $z \sim 0.55$.

\section {Probing SHELS}
\label{shelsprobes}
To evaluate the correspondence between systems in the redshift survey and 
significant peaks in the convergence map, we develop an approach for sampling
the redshift survey in narrow cones. We choose
the opening angle of the cone with attention to the resolution of the weak
lensing map, the density of the redshift survey in the relevant 
redshift range, the limitations of a magnitude limited redshift survey, the properties of massive clusters, and previous
evaluations of convergence maps for cluster identification.

Our goal is to assess the {\it efficiency} and {\it completeness} of the convergence map in the identification of massive cluster halos. We define {\it efficiency} as the
fraction of convergence map peaks with $\nu \geq3.5$ that correspond to a SHELS cluster with a line-of-sight velocity dispersion above the solid $\nu=3.5$ threshold curve of Figure \ref{fig:sensitivity_band.ps}. In other words, the efficiency is the fraction of weak lensing peaks that correspond to appropriately
massive systems. We define {\it completeness} as the fraction of all clusters in SHELS with
line-of-sight velocity dispersions above the threshold curve that also correspond to weak lensing peaks
with $\nu \geq3.5$. In other words, the completeness is the fraction of appropriately massive systems in SHELS that are detected as weak lensing peaks. These definitions exclude superpositions from consideration. They also exclude coincidences between the abundant low velocity dispersion systems in the redshift survey with convergence peaks. Our definitions are similar to those adopted by Hamana et al. (2004) in their simulations of the
efficacy and completeness of weak lensing surveys for massive cluster identification.

In this section, we show that sampling the redshift
survey in narrow cones recovers the four known x-ray clusters 
with rest frame line-of-sight velocity dispersion $\gtrsim 600$ km s$^{-1}$ (Section \ref{probes}). We then use similar probes through SHELS along the lines-of-sight toward the 12 DLS convergence peaks with $\nu \geq 3.5$  and examine the presence/absence of obvious massive clusters (Section \ref{shelscandidates}). This exercise gives us a first estimate of the {\it efficiency} of the convergence map for cluster detection. From SHELS we can compute a velocity dispersion for candidate systems along the line-of sight toward the most significant convergence peaks thus refining our estimate of efficiency. We also extend the investigation of the correspondence between SHELS systems and convergence peaks to $\nu \gtrsim 1$ to make a first assay of the presence of signal in the convergence map at low $\nu$ and uncover the expected increasing fraction of noise peaks with decreasing significance in the convergence map (Section \ref{vdisps}). X-ray observations cover only a small fraction of F2; Section \ref{xraysF2} discusses the SHELS velocity dispersions and the DLS detection/non-detection of the seven known extended cluster x-ray sources. Finally, we use SHELS as a basis for evaluating  the {\it completeness} of the convergence map by comparing the DLS weak lensing detections with the number of clusters in SHELS that should be detected. Section \ref{shelsclusters} discusses the clusters in SHELS that do not appear as $v \geq 3.5$
peaks in the convergence map even though their optical properties indicate that they should produce a 
peak in the map.

\subsection {Using SHELS to Evaluate Convergence Map Peaks}
\label{probes}

Schirmer et al. (2007) and Gavazzi \& Soucail (2007) develop uniform
procedures for evaluating the coincidence of concentrations of galaxies 
with convergence map peaks. Schirmer et al. (2007) compare 
the galaxy counts with R $\sim$ 17---22  in a
cone of 2$^{\prime}$ radius centered on a lensing peak with
the galaxy density in the surrounding field. They assign significance classes
to the excess counts based on Poisson statistics. Gavazzi \& Soucail use
the distribution of photometric redshifts for galaxies brighter than 
i$^{\prime}$ = 23 in
2$^{\prime}$ radius probes. They 
subtract a background photometric redshift distribution from 
each of the probes. They then compute the 
background from the survey area more than
6$^{\prime}$ from any peak. 

The approaches of both Schirmer et al. (2007) and Gavazzi \& Soucail (2007)
are based on complete photometric surveys of their fields. They evaluate 
the significance of cluster candidates internal to their surveys.  

We follow a similar approach with a complete redshift survey. Because the
redshift survey is complete we can sample the survey itself to evaluate the
significance of galaxy condensations along the line-of-sight to lensing peaks.
The redshift survey is shallower than the corresponding photometric surveys.
We demonstrate below that redshifts for the more luminous galaxies 
in condensations at 
redshifts from 0.2--0.53 are  adequate to 
demonstrate correspondence (or lack of it) between the weak lensing peaks and 
clusters of galaxies in the redshift survey. 
We briefly explore the dilution of 
a ``cluster'' signal in photometric redshift data in Section \ref{photoz}. The
sensitivity plot (Figure \ref{fig:sensitivity_band.ps}) 
suggests that we need not look deeper than
$z \sim 0.55$ to identify clusters corresponding to the weak lensing peaks;
more distant systems must be very rich and they should be obvious in the 
photometric data. Kubo et al. (2009) find no evidence for such
higher redshift systems.

Probes through the redshift survey centered on each galaxy in the survey
contain all of the
selection effects which impact a probe toward a lensing peak. Thus comparison
of these probes with similar probes toward the peaks should provide a 
robust measure of the significance of cluster candidates in the redshift
survey. We have demonstrated by experiment that  centering of probes on 
galaxies or on random points in the region has no effect on the results of
our analysis.

The steps in our candidate cluster identification procedure are:

\begin{enumerate}

\item We sample the SHELS survey complete to R = 20.3 in  
{\it test cones} 
with  3$^{\prime}$ radius centered on each galaxy 
in the complete redshift survey. 
The cones are large enough to detect a galaxy cluster
across the redshift range we sample.

\item In each {\it test cone} we count 
the number of additional galaxies, N$_{gal}$, within
a bin of $1600 (1 + z)$ km s$^{-1}$ centered on the survey galaxy. The 
bin size is comparable with the extent in redshift space of 
systems we wish to identify. The bin width is conservatively large.

\item In each of the redshift bins of item 2, 
we evaluate the mean occupation and the
variance ($\sigma_{SH}$) across the entire survey. 
We then identify the set of {\it test cones} at each 
redshift which are 5$\sigma_{SH}$ above the mean occupation and contain at least
6 galaxies (N$_{gal} \geq 5$). The minimum number 
of galaxies enables computation of a
dispersion (albeit with large error). The 5$\sigma_{SH}$ limit restricts the
sample to high peaks which are reasonable candidate clusters especially at
the peak sensitivity of the survey, $z \sim 0.3$.
Both the 6 galaxy and 5$\sigma_{SH}$ limits are
generous; they admit many peaks well below the expected detection 
threshold for the weak lensing map (Figure \ref{fig:sensitivity_band.ps}). We call these probes 
{\it 5$\sigma_{SH}$ probes} hereafter.

\end {enumerate}

Figure \ref{fig:5sigmamap.ps} shows a map of the F2 region. Each gray point marks 
the position of a 
galaxy  at the center of a $5\sigma_{SH}$  probe.
As a basic test of the efficacy of these probes in identifying known 
clusters of galaxies, the open circles in Figure \ref{fig:5sigmamap.ps} show that
the locations of    
four known x-ray clusters, XMMU J091935+303155, CXOU J092026302938, CXOU J092053+302900, and CXOU J092110+302751, coincide with highly ranked probes.
The cluster XMMU J091935+303155 is undersampled in the redshift survey because a 
saturated bright star (R$\lesssim 18$)  is superposed near its center. 
We comment further
on these clusters in Section \ref{xraysF2}.

Kubo et al. (2009) find 12
peaks in the convergence map with signal-to-noise $\nu \geq 3.5 $. The
numbers in Figure \ref{fig:5sigmamap.ps} are centered on the coordinates of these 
significant weak lensing peaks. Table 1
of Kubo et al. (2009) lists the
coordinates and value of $\nu$ for these peaks. The 
significance of these peaks is similar to the peaks considered in other
convergence maps and in the simulations by Hamana et al. (2004).

In Figure \ref{fig:5sigmamap.ps} the 
two most significant weak lensing peaks are coincident 
with the two x-ray clusters, CXOU J092026+302938 and CXOU J092053+302900.
These two clusters coincide with the cluster A781.
All of the weak lensing peaks except 4, 6, 8 and 12 lie 
within 3${^\prime}$ of the center
of at least one well-populated probe. 

There are obviously many well populated
probes without any associated significant weak lensing peak. 
Many of the well-populated SHELS probes correspond to groups 
at a redshift $\lesssim 0.3$. From Figure \ref{fig:sensitivity_band.ps} we would not expect these systems with $\sigma_{los} \lesssim 500$ km s$^{-1}$ to produce a significant signal in the $\kappa$ map.

To further assess the meaning of the weak lensing peaks we 
examine the redshift distributions along the line-of-sight 
toward each of the 
significant peaks. We ask which lines-of-sight intersect a
cluster with a velocity dispersion large enough to plausibly account for 
the weak lensing peak (Section \ref{vdisps}). We also investigate the coincidence of SHELS systems with
lower significance convergence peaks with 1$ < \nu < 3.5$ (Section \ref{vdisps}). Finally, we  check all of the 5$\sigma_{SH}$ peaks 
in Figure \ref{fig:5sigmamap.ps} to see whether any others contain systems with large enough 
velocity dispersion that they should be detected according to the sensitivity 
plot (Figure \ref{fig:sensitivity_band.ps}; Section \ref{shelsclusters}). We can then evaluate the completeness of the DLS cluster detections (see Sections \ref{weakcluster} and \ref {discussion}).

\subsection {Convergence Map Peaks and Candidate Galaxy Systems}
\label{shelscandidates}
In this section we evaluate candidate systems along the line-of-sight toward 
weak lensing peaks. We begin with the 12 most significant DLS peaks. We 
extend the discussion to all DLS peaks with $\nu \gtrsim 1$ in
Section \ref{shelsclusters}. 

Figures \ref{fig:PeakHisto.1-3.ps}, \ref{fig:PeakHisto.4-6.ps}, \ref {fig:PeakHisto.7-9.ps},
and \ref{fig:PeakHisto.10-12.ps} show the redshift, $z$,  distributions of galaxies  within
3$^{\prime}$ probes centered on each of the 12 highest
significance DLS peaks (see Figure \ref{fig:5sigmamap.ps}). 
The bins are 800(1 + z) km s$^{-1}$. The dark histogram shows 
galaxies with R $\leq 20.3$; the gray histogram shows galaxies 
with R $\leq 20.6$. The histograms
fall in three categories: (1) there is an obvious single peak which probably 
corresponds to a system of galaxies  (2) there is more than one apparently
significant peak, and 
(3) there are no significant peaks. 

The probes toward peaks 1, 2, and 5 each contain  
an impressive peak. These peaks correspond to
clusters of galaxies; we evaluate the dispersions in Section \ref{vdisps}. 

Peak 3 contains several well-populated  peaks at redshifts 0.28 --- 0.34.
Peaks 4, 6, 8, and 12 contain no well-populated peaks. If there 
are condensations
in redshift space near these peaks, most of the galaxies are more than 
3$^{\prime}$ from the position of the weak lensing peak. With a larger search radius of 4$^{\prime}$, Kubo et al. (2009) identify candidate systems associated with convergence map peaks 8 and 12.

Figure \ref{fig:4panel5vs3.ps} shows  higher resolution histograms for weak lensing peaks 
3 and 5 in both 3$^{\prime}$ and 6$^{\prime}$ probes. The upper histograms show the distributions for the magnitude limited
samples with R $\leq 20.3$; the lower histograms show all 
of the galaxies with R $\leq$ 20.6.
The difference in the redshift distributions is obvious particularly in the
better sampled histograms. Peak 5 is a cluster with a rest frame velocity 
dispersion of 729$\pm$41 km s$^{-1}$ (3$^{\prime}$ probe); the line of sight toward peak 3 contains a superposition
of three peaks with rest  frame velocity dispersions of 150, 260, and 340 km s$^{-1}$

A detailed model is necessary to assess whether the 
superposed low mass systems along the line-of-sight toward peak 3 
can account for the weak 
lensing signal. This detailed discussion is beyond the scope of this paper;
for the rest of the discussion we focus on lines-of-sight which include 
a plausibly rich cluster. We concentrate on evaluating the 
weak lensing map as a method of identifying these rich systems. 

The lines-of-sight toward peaks 7 and 9 each contain a peak populated by 
$\sim$ 10 galaxies. Peak 10 has a peak at 0.185  along with a peak
at 0.53. The redshift survey is sparse at z$\sim$ 0.5; thus the number 
of galaxies in the peak at 0.53 is still indicative of the presence of a 
system (see Section \ref{vdisps}). The fainter 
galaxies obviously enhance the peak. The situation is similar for the
line-of-sight toward peak 11; there is a peak at z = 0.53 again enhanced by
the fainter galaxies (see Section \ref{xraysF2}).  

The apparent systems which appear as peaks in the redshift survey have a large 
range in velocity dispersion. We next determine which of these peaks 
correspond to systems with velocity dispersion (mass) sufficient 
to account for the weak lensing signal.  

\subsection{Velocity Dispersions of Weak Lensing/SHELS Candidate Systems}
\label{vdisps}
In this section we compute velocity dispersions for systems within our 3$^\prime$ probes. To
construct a complete catalog of massive systems which should be detected by the DLS at $\nu \geq 3.5$, we consider
all apparent systems with rest frame line-of sight velocity dispersion $\sigma_{los,3} \gtrsim 500$ km s$^{-1}$ measured from the R$\leq$ 20.3 catalog. The velocity dispersion limit lies significantly below the $v = 3.5$ threshold in Figure 
\ref {fig:sensitivity_band.ps}. We also examine weak lensing peaks with statistical significance
$\nu \gtrsim 1$  
to examine the overall signal in the weak lensing map and to 
evaluate the frequency of accidental superpositions between weak lensing 
peaks and clusters in SHELS.

We use SHELS redshifts for galaxies with R $\leq$ 20.6 to compute the rest frame velocity 
dispersion, $\sigma_{los,3}$,  for each of the candidate systems within a 3${^\prime}$
probe centered on the position of the weak lensing peak. We also 
compute the rest frame velocity dispersion for each candidate system within
a  6${^\prime}$ probe concentric with the 3${^\prime}$ probe. We make this comparison
because the determination of the velocity dispersion of the clusters which comprise the A781 complex
is problematic as a result of system overlap in redshift space. In general, the comparison 
of these two apertures also gives a measure of the systematic error in the determination of the velocity 
dispersion.

Table \ref{tbl:VDisp1} lists the candidate systems and their properties
for both a 3${^\prime}$ and  a 6${^\prime}$  radius centered
at the listed RA$_{2000}$ (column 4) and DEC$_{2000}$ (column 5). The Table includes systems along the line-of sight toward weak lensing peaks with $\nu > 1$ as well as candidate SHELS systems where there
is no corresponding peak in the convergence map.
Table \ref{tbl:VDisp1} lists the candidate cluster name (column 1), the rank of the corresponding DLS peak if there is one (column 2), the $\nu$ of the weak lensing peak (column 3), the mean redshift of the SHELS candidate system (column 5), the rest frame line-of-sight velocity dispersion within the 3$^{\prime}$ radius, $\sigma_{los,3}$ (column 7), the number of survey galaxies in the 3$^{\prime}$ system, N$_3$ (column 8), the bootstrap error in $\sigma_{los,3}$, err$_3$ (column 9), the  rest frame line-of-sight velocity dispersion within the 6$^{\prime}$ radius, $\sigma_{los,6}$ (column 10), the number of survey galaxies in the 6$^{\prime}$ system, N$_6$ (column 11), and the bootstrap error in $\sigma_{los,6}$, err$_6$ (column 12).
For weak lensing peaks 1 and 2,the Table lists only 6$^\prime$ quantities. We discuss this issue in detail below. In all other cases we limit the the range where we compute the velocity dispersion
with standard 3$\sigma$ clipping. For nearly all of the cluster candidates (regardless of
a counterpart in the weak lensing map),  the rest frame
velocity dispersions in the two apertures differ 
by $\lesssim$ 2$\sigma$. 
In fact, the only exception is DLS peak 15 where the difference is 
2.2 $\sigma$.

The abundance of systems of galaxies increases steeply with decreasing line-of-sight velocity dispersion.
We limit our discussion of candidate lensing systems to massive halos with $\sigma_{los,3} \gtrsim 500$ km s$^{-1}$. Coincidences of weak lensing peaks with individual much less massive systems are common, as are
superpositions along the line-of-sight. Of course, the greater abundance of these systems implies a larger
chance of mismatching halos with weak lensing peaks (see Hamana et al. (2004)).  Analysis of this issue is beyond the scope of this paper.

For the range of candidate cluster redshifts in Table \ref{tbl:VDisp1}, 
the 3$^{\prime}$ radius corresponds to 0.38 Mpc (z = 0.12) and 
to 1.1  Mpc (z = 0.53).
These radii are within r$_{200}$, the radius where the enclosed average mass density, $\rho(<r)_{200} = 200
\rho_c$. Here $\rho_c$ is the critical density. The radius $r_{200}$, a proxy for the virial radius, ranges from 1.0 to 2.0 Mpc 
for well-sampled clusters with
rest frame line-of-sight velocity dispersions in the
range 500 km s$^{-1}$ to 1000 km s$^{-1}$ (Rines et al. 2003; Rines \& Diaferio (2006). According to the
scaling relations of Rines \& Diaferio (2006), the corresponding range of masses, M$_{200}$, within r$_{200}$ is 1.2$\times 10^{14}$ M$_\odot$ to 1.2$\times 10^{15}$ M$_\odot$. 

Rines \& Diaferio (2006)
show that the velocity dispersion in a cluster decreases with radius. For 
clusters at larger redshift the tendency to underestimate the velocity 
dispersion in our physically larger aperture is compensated by the increasing
difficulty of eliminating velocity outliers in the sparser samples. 

Determination of the central velocity dispersion for weak lensing peaks 1 and 2 is
problematic because they overlap substantially in redshift space. Geller et 
al. (2005) published rest frame velocity dispersions of 741$^{+35}_{-40}$ km s $^{-1}$ 
and 674$^{43}_{-52}$ km s$^{-1}$ 
for peaks 1 and 2,  respectively, based on identification of cluster members in a 
friends-of-friends group finding algorithm applied to less complete SHELS 
data. Here we explored the region of the A781 complex by first assessing the
global redshift distribution within a 30$^{\prime}$ region centered midway
between the coordinates listed in Table \ref{tbl:VDisp1} for these systems. There are
two well-defined peaks centered at redshifts 0.2915 (peak 1) and 0.3004 (peak 2) in essential
agreement with the mean redshifts for these systems in Geller et al. (2005). 
Application of the routine {\it mrqmin} (Press et
al. 1992) returns velocity dispersions for peaks 1 and 2 that substantially underestimate the central
velocity dispersions because they sample well into the cluster infall region 
(Rines et al.2003; Rines \& Diaferio 2006).  The mean redshifts are, however, robust and we use them
to compute the velocity dispersion in the central regions of the clusters.

To estimate the central rest frame velocity dispersion of peaks 1 and 2,
we fix the mean redshift at the values obtained for the 
global sample (Table \ref{tbl:VDisp1}).
For samples within a 3$^{\prime}$ radius of each center, a fit of two Gaussians
in the Press et al. (1992) routines does not converge. Within a 6$^{\prime}$ radius
(a physical radius of $\sim$ 1.5 Mpc, roughly the expected r$_{200}$ for these clusters)
around each of the two centers, the fits converge to the rest frame dispersions
in Table \ref{tbl:VDisp1}. The error in Table \ref{tbl:VDisp1} is an approximate and conservative estimate;
the errors are dominated by systematic issues in separating the two clusters 
in redshift space. This approach yields systematically larger velocity dispersions than the
group-finding algorithm used in Geller et al. (2005), but the differences are within the large error. Regardless of the method of assessing the velocity dispersion of these clusters, they both exceed the 
$\nu = 3.5$ threshold 
(Figure {\ref {fig:sensitivity_band.ps}).

There are 10 SHELS candidate systems along the line-of-sight toward weak lensing peaks with $\nu > 1$. In
one case, weak lensing peak 10, there are two systems along the line-of sight, one at 
$z = 0.18$ along with a more  massive system at $z = 0.53$. The median separation between the 
lensing peak positions and the centers of the candidate galaxy cluster is 0.5$^\prime$ in agreement with the similarly small separations found in previous surveys (Gavazzi \& Soucail 2007; Schirmer et al. 2007). 
Figure \ref{fig:ranksigma.ps} shows the DLS weak lensing peak rank as a function of statistical significance. 

Solid dots in Figure \ref{fig:ranksigma.ps} represent cases where SHELS
reveals a cluster along the 
line-of-sight to the peak with a rest frame velocity dispersion $\gtrsim 500$
km s$^{-1}$ and $z < 0.55$.
It is interesting to note that, as expected,  the fraction of ``cluster
detections'' (solid dots) is larger for $\nu_{DLS} \geq 3.5$ than at lower significance.. In Section \ref{weakcluster} we estimate the accidental coincidence rate and show that there is some signal in the weak lensing map at $\nu_{DLS} < 3.5$ even though, as expected, it is diluted by an increasing abundance of noise peaks as $\nu_{DLS}$ decreases.  In contrast with Kubo et al. (2009), we find only 4 
(rather than 10) convincing coincidences between rich clusters and weak lensing peaks with $\nu \geq 3.5$. The main reasons
for the difference between our results and Kubo et al. (2009) are (1) our use of the redshift survey to compute
line-of-sight velocity dispersions and (2) our use of a smaller search radius (consistent with the small offsets between lensing peak positions and galaxy system centers observed by other investigators from both data and simultions). Kubo et al. (2009) count superpositions like peak 3 as weak lensing
detections; we count only candidate systems which have a large enough SHELS line-of-sight velocity dispersion to exceed the
$\nu = 3.5$ threshold.  We conclude that the weak lensing detection efficiency for $\nu \geq 3.5$ is 33\%.

\subsection {Known X-ray Clusters in the DLS Field}

\label{xraysF2}
Chandra and XMM observations cover 14\%
of the DLS field with widely variable sensitivity. As a result of these observations there
are seven known extended cluster x-ray sources in the field (Figure \ref{fig:comparison_map.ps}). 
Both Chandra and XMM
observations target the region near A781 which, 
remarkably, contains 4 cluster x-ray sources (Table 
\ref{tbl:Xrays} and Figure \ref{fig:comparison_map.ps}). In order of right ascension, Table \ref{tbl:Xrays} lists the x-ray cluster name (column1), the rank of any corresponding DLS peak (column 2), the RA$_{2000}$ and DEC$_{2000}$ (columns
3 and 4), the rest frame line-of-sight velocity dispersion determined from the SHELS galaxies in the system with R $< 20.6$ in a 3$^{\prime}$ cone (except for  CXOU J092026+302938 and CXOU J092053+302900 where
we quote the 6$^\prime$ dispersions from Table \ref{tbl:VDisp1}) centered on the x-ray position (column 5), the number of galaxies in the system (column 6) the bootstrap error in the velocity 
dispersion (column 7), and the {\it Chandra} x-ray flux from the literature (column 8). Double line entries for a single x-ray system indicate superposed structure in SHELS along the line-of-sight. 

The two clusters CXOU J092026+302938 and CXOU J092053+302900
are  at  redshifts of 0.3004 and 0.2915 respectively; 
these two clusters correspond to the original A781. They are also responsible
for the two most significant weak lensing peaks in the Kubo et al. (2009) map.
The cluster CXOU J092026+302938  is complex; CXOU J092053+302900 is a simpler system
(Wittman et al. 2006; Sehgal et al. 2008).

Khiabanian \& Dell'Antonio (2008)  developed the lensing map 
reconstruction technique applied by Kubo et al. (2009). In addition to the
map with a uniform 1.5$^{\prime}$ resolution, Khiabanian \& Dell' Antonio (2008)
show a map with a resolution of 0.9$^{\prime}$ in the regions around A781. 
This map reveals a detection of a third x-ray cluster CXOU J092026+302938 and CXOU J092053+302900 with a S/N$\sim$ 5.7.
SHELS shows that
this cluster has a mean redshift of 0.43 and a velocity dispersion of 
754 $\pm$ 92 km s$^{-1}$ (Table \ref{tbl:Xrays}).

The fourth extended cluster x-ray source in the region of A781, XMMU J091935+303155,
is a puzzling and complex case. There is a well-defined, blue arc
associated with the first-ranked galaxy located at RA$_{2000}$ = 9:19:35.063, DEC$_{2000}$ = 30:31:56.627, very close to the xray center. Our attempt to measure a
redshift for the arc with the Blue Channel spectrograph on the MMT
resulted in detection of a blue continuum but no emission lines.
Thus the redshift of the arc is probably $\gtrsim 1.2$. The mean
cluster redshift, 0.43, is essentially the same as the redshift of the first-ranked 
galaxy (0.4276). This cluster is undersampled in SHELS because a region
very close to the cluster center is excised around saturated bright star (R$\lesssim 18$).

Although the velocity dispersion and redshift of XMMU J091935+303155
should put it above the 
detection threshold for the weak lensing map, there is no significant peak
associated with this system. The XMM observations (Table 6 of Sehgal et al. (2008)) show that
XMMU J091935+303155 has about twice the x-ray flux and nearly the same temperature as
CXOU J092110+302751. CXOU J092110+302751 is at the same redshift and is cleanly detected in the higher resolution DLS weak lensing map (Khiabanian \& Dell'Antonio 2008).
In contrast, we find only a $\nu = 0.5$
enhancement in the convergence map at the position of XMMU J091935+303155. Sehgal et al. (2008) make an 
{\it a posteriori} model fit to 
the convergence map based on the x-ray data and argue for a weak detection. Both the SHELS
data and the XMM observations suggest that the cluster should appear at a $\nu > 3.5$ in the convergence map.

In separate deep imaging with Chandra, Wittman et al. (2006) discovered 
three more extended x-ray sources: CXOU J091551+293637, CXOU J091554+293316,
and CXOU J091601 +292750 (see Table \ref{tbl:Xrays} and Figure \ref{fig:comparison_map.ps}). In Figure \ref{fig:5sigmamap.ps},
possible optical counterparts of CXOU J091551+293637 and  CXOU J091554+293316 appear
as populated {5$\sigma_{SH}$} probes. CXOU J091551+293637 and  CXOU J091554+293316 are adjacent to weak lensing peak 11 and 8 respectively.
These systems do not appear in Table \ref{tbl:VDisp1} because the velocity dispersion for samples centered on the weak lensing peaks and/or for samples limited to R $< 20.3$ fail our minimum cut of 500 km s$^{-1}$.

Table \ref{tbl:Xrays} lists line-of-sight velocity dispersions for condensations of galaxies along the line-of-sight toward these three extended x-ray sources. 
For CXOU J091551+293637, the probable first-ranked  galaxy in the x-ray emitting system is within 30$^{\prime\prime}$ of the x-ray center and has a redshift of 0.53. We thus identify the extended x-ray source as a rich group at $z = 0.53$ with a rest frame line-of-sight velocity dispersion of 360$\pm$70 km s $^{-1}$. The first-ranked galaxy and the ten other probable members of the system are all fainter than R = 20.3. Thus foreground structure at z $\sim$ 0.18 is responsible for the 5$\sigma_{SH}$ probes in Figure \ref{fig:5sigmamap.ps}. In Table \ref{tbl:Xrays} we quote a velocity dispersion for this
foreground structure that is merely the spread of redshifts within the large-scale structure intercepted by our probe.  

The x-ray source CXOU J091554+293316 has the lowest flux of the three (Wittman et al. 2006). There is no convincing optical counterpart; the nearest galaxy is an arcminute away from the x-ray center and only 7 other galaxies within our 3$^{\prime}$ probe appear clustered at the $z = 0.18$;
a normal x-ray emitting group at this redshift should be easily detectable  with many members in a deep survey like SHELS. We list an apparent velocity dispersion in Table \ref{tbl:Xrays}, but this dispersion is again a measure of the spread of redshifts within the large-scale structure intercepted by our probe.

The center of the southernmost extended x-ray source, CXOU J091601+292750, is within 10$^{\prime\prime}$ of a galaxy at $z = 0.53$. An additional six galaxies at similar redshift yield an estimate of the rest frame line-of-sight velocity dispersion for this system of 523$\pm$71 km s$^{-1}$. The foreground structure seen in the lines of sight toward  CXOU J091551+293637 and  CXOU J091554+293316 is also present here. Again the velocity dispersion for this structure at $z = 0.18$ is indicative of the 
spread of redshifts within the foreground large-scale structure. There is no weak lensing peak near this x-ray system. 

The rest frame velocity dispersions for all three systems (CXOU J091551+293637, CXOU J091554+293316,
and CXOU J091601+292750) should
place them below the $\nu = 3.5$ detection threshold for the weak lensing map 
(Figure \ref{fig:sensitivity_band.ps}). Kubo et al. (2009), however, associate these clusters with
weak lensing peaks 8 and 11. It is conceivable that foreground structure at $z = 0.18$
and overlap among the clusters boosts the weak lensing
signal; detailed modeling beyond the scope of this paper 
is necessary to understand
whether these superpositions are adequate to explain the apparent 
weak lensing detection.

Among the seven known x-ray clusters in the DLS fields, only three (CXOU J092026+302938, CXOU J092053+302900 and CXOU J092110+302751) are cleanly
detected as peaks with $\nu > 3.5$ in the convergence map provided the resolution 
is adequate. The cluster XMMU J091935+303155, detected by XMM and SHELS but not by the DLS, is puzzling.
Another 3 clusters at greater redshift (and lower rest frame velocity 
dispersion)  may be 
detected in the DLS map as a complex superposition; the SHELS velocity dispersions and the
sensitivity curve for the convergence map suggest that none of these systems should be detected on their own. 

This analysis of 
x-ray observations of a small portion of the 
DLS field demonstrates some problematic issues for construction of
catalogs of clusters of galaxies from weak lensing maps: (1) coarse 
resolution of the weak lensing map dictated by the density of sources 
can reduce the number of clusters in an uncontrolled way, (2) superpositions
may increase the number of apparent detections in a complex way, and (3)
inconsistencies among different cluster detection methods need to be understood both from
simulations and from larger observational samples where direct comparison is possible. Berg\'e et al. (2008) are, for example, undertaking such an observational project in the x-ray.

\subsection {SHELS Cluster Candidates}
\label{shelsclusters}
SHELS obviously enables identification of cluster candidates independent of the
weak lensing map. There are many reasonable approaches to the identification
of candidates. For consistency, we use the 3$^{\prime}$ probes of 
Figure \ref{fig:5sigmamap.ps}. Other algorithms including the identification of systems
with a friends-of-friends algorithm yields very similar results for these 
dense, well-populated systems (Ramella et al. 2010).

To construct a catalog of cluster candidates for z $< 0.55$ we start 
with the probes shown in the map of Figure \ref{fig:5sigmamap.ps}. We first identify 
all probes with at least 6 galaxies with R$\leq$ 20.3
(significance $\gtrsim 5 \sigma_{SH}$) in any redshift 
interval, 1600(1 + z) km s$^{-1}$, and
with apparent velocity dispersion $\gtrsim 500$ km s$^{-1}$. 
We then examine all overlapping probes and choose the most populated
to represent the system. We choose the center of this probe as the indicative center of the
system.

The probes in Figure \ref{fig:5sigmamap.ps} are based solely 
on galaxies with R $<$ 20.3. We
next include the additional galaxies with 20.3 $<$ R $\leq 20.6$. The broad window we use to construct
the map may admit several groups with small velocity dispersion as in Peak 3.
Increased sampling enables elimination of several systems where we can identify better defined multiple peaks  as convincing
superpositions. We list all of the SHELS cluster candidates in Table \ref {tbl:VDisp1}.  For completeness we include all of the candidate clusters
we find with line-of-sight velocity dispersion $\gtrsim$ 500 km s$^{-1}$ even though many of these should not be detected at high significance in the DLS weak lensing map. For redshifts $\lesssim 0.4$ we 
easily identify clusters with line-of-sight velocity dispersion $\sim$ 500 km s$^{-1}$; for larger redshift the lowest velocity dispersion system has $\sigma_{los,6} = 642+/-70$ km s$^{-1}$. 

The sample of SHELS candidate clusters with  
velocity dispersions large enough to be detected in the DLS weak lensing map should be complete (see Figure \ref{fig:sensitivity_band.ps}).
The set of 20 SHELS clusters (Table \ref{tbl:VDisp1}) includes 10  systems along the line-of-sight
toward weak lensing peaks with $\nu \gtrsim 1$.  We also find the two additional x-ray
clusters SHELS J0921.2+3028 (CXOU J092110+302751) and SHELS J0919.6+3032 (XMMU J091935+303155). A higher resolution weak lensing map detects the first of these; there is no weak lensing detection of the second
(see Section \ref{xraysF2}). Considering the errors in the line-of sight velocity dispersion, $\sigma_{los,6}$, the SHELS candidate cluster list  includes eight clusters with line-of-sight velocity dispersion large enough that they should be detected with $\nu \gtrsim 3.5$ in the DLS map at the resolution we explore here; only four of them are detected (we exclude
J0921.2+3028 which requires a higher resolution map for DLS detection). All four of the
undetetected clusters are at a redshift $z \sim$ 0.4. 
It is interesting that three of these clusters (two at $z \sim$ 0.4) are along the line-of-sight toward DLS peaks with $1.6 \leq \nu < 2.9$. Details of the cluster structure, voids along the line-of sight, and noise in the $\kappa$ map may all contribute to reduction of the weak lensing signal (Hamana et al. 2004). Our data do not allow discrimination among these possibilities.

\section {Weak Lensing and  Cluster Detection}
\label{weakcluster}

Figure \ref {fig:sigmavsz.ps}  summarizes the joint detection of systems
in the DLS and in SHELS. Each of the symbols represent one of  the 20 SHELS candidate clusters
in Table \ref{tbl:VDisp1}.
The solid curve shows the expected $\nu = 3.5$ detection threshold in the DLS.
Figure \ref{fig:sigmavsz.ps} indicates sensitivities of $\nu =3 $ (lower dotted curve) and $\nu = 4$  (upper dotted curve).
We use $\sigma_{los,6}$ for uniformity across the SHELS sample.

The four filled triangles indicate the SHELS clusters coincident with
DLS $\nu \geq 3.5$ peaks. Note that clusters at
z = 0.1844 and z = 0.5343 are superposed along the line-of-sight toward DLS peak 10; we show the
superposed lower redshift system as an open triangle.
Among the top 12 DLS peaks there are 4 detections of clusters of galaxies
(peaks 1, 2, 5 and 10). Peaks 3, 8, and 11 may reflect lensing signal from
superpositions (see Sections \ref{shelscandidates} and \ref{xraysF2}). Conservatively, the efficiency of cluster
detection is 33\% for weak $\nu \ge 3.5$ weak lensing peaks. The efficiency 
for the 7 $ \nu \geq 4$ peaks increases to 43\%.

The filled star in Figure \ref{fig:sigmavsz.ps} corresponds to the x-ray cluster SHELS J0916.2+2949
CXOU J092110+302751. 
In the regions around A781, Khiabanian \& Dell'Antonio (2008) construct a higher resolution map which 
has a $\nu \sim 5.7$ peak coincident with this cluster. Wittman et al. (2006) also identify it in their
higher resolution weak lensing map of F2. 

The five open squares correspond to weak lensing peaks with 1 $\leq \nu < 3.5$ and 
with a SHELS cluster along the line-of-sight.  If the weak lensing 
peaks were uncorrelated with systems of galaxies we would expect 
2.5$\pm$0.5  of the 63 weak lensing peaks would fall within the  20 3$^\prime$ probes where we identify
clusters in SHELS. This estimate of chance coincidences is probably overly generous because the median separation
between the center of a SHELS cluster and a weak lensing peak is only 0.5$^\prime$ with a maximum separation
of 1.5$^\prime$. 
If we require that the centers of the weak lensing peaks and the SHELS clusters be less than $1.5^\prime$, the expected number of accidental coincidences obviously drops by a factor of 4 to 0.6$\pm$0.1,
a negligible contamination.  We conclude that there is  signal in the weak 
peaks at $\nu < 3.5$; the coincidences between SHELS clusters and lower significance DLS lensing peaks are probably not accidental. As expected the efficiency of halo detection decreases with the significance of the weak lensing peaks. Of course, the clusters in the DLS field are embedded in the large-scale structure and contribute to the cross-correlation signal Geller at al. (2005) find between the DLS convergence map and a
velocity dispersion map constructed from SHELS. 

The 9 open circles  are candidate clusters in SHELS with no corresponding DLS
peak at significance $\nu \gtrsim 1$. Again the five open squares represent candidate SHELS clusters along the line-of-sight toward DLS peaks with significance 1$\leq \nu < 3.5$. Among these systems indicated by open squares and circles, four should be 
detected by the DLS at $\nu \geq 3.5$; two of these systems (open squares) are along the line-of-sight toward lower significance 
weak lensing peaks. Thus the completeness of the DLS cluster detection (the fraction of SHELS clusters with velocity dispersions
above the $\nu = 3.5$ threshold detected by the DLS) is 
$\sim 50$\%. This completeness level is similar to the one predicted by the simulations of Hamana et al. (2004): we discuss this similarity further  in Section \ref{discussion}. 

Figures 12, 13, and 14 show DLS images of the central $6^\prime \times 6^{\prime}$ of each of the nine clusters in SHELS with rest frame line-of-sight velocity dispersion measurements placing them about the DLS $\nu = 3.5$ threshold. Among these clusters  
SHELS J0920.9+3029 and  SHELS J0920.4+3030 (they are coincident with A781 at z$\sim 0.3$ ) and SHELS J0916.2+2949 at $z \sim 0.5$ are obviously very rich systems. Among the five clusters at $z \sim$ 0.4, there is no obvious  optical property that should discriminate between weak lensing detection and non-detection. 
At $z \sim0.3$, the peak sensitivity of both SHELS and the DLS,  all three clusters within SHELS which should be detected in the DLS map  at $\nu \geq 3.5 $ correspond to  
significant weak lensing peaks. The situation at greater redshift is puzzling. Clusters detected as weak lensing peaks and those undetected have the same rest frame line-of-sight velocity dispersions within the errors. For the clusters at $z \sim 0.4$ the number of galaxies within SHELS for the detected cluster
is about twice the number in the undetected ones.  
These variations in population are unlikely to explain the weak lensing non-detections; variations of a factor of two at a fixed cluster mass (velocity dispersion) are typical of the cluster population at lower redshift (see e.g. Lin et al. 2004).

Even for peaks in the weak lensing map with significance $\nu \geq 3.5 $, cluster detection is neither 
complete nor efficient (Hamana et al (2004); Hennawi \& Spergel (2005)). The weak lensing map is $\sim$ 50\% complete relative to the catalog derived from the SHELS survey. Among the 12 DLS peaks with high significance, only 33\% are associated with
massive clusters. Admittedly the sample of significant weak lensing peaks is small, but the comparison
with SHELS suggests that observational assessments of weak lensing maps based on counts along the line-of-sight and/or on photometric redshifts may lead to overly optimistic assessments of weak lensing as a tool for the construction of catalogs of massive systems in the universe.

\section {Degrading SHELS Redshifts --- A Perfect Photo$z$ Model} 

\label{photoz}

Previous techniques for identifying candidate systems corresponding to
convergence map peaks include the use of galaxy counts, photometric
redshifts, and x-ray images. In the largest datasets to date, the efficiency of cluster detection ranges from 45\% (Schirmer et al. 2007) to 80\% (Miyazaki et al.  2007).
 
In this section we focus on the use of
photometric redshifts and demonstrate by example that this approach can easily fail to discriminate
between a superposition along the line-of-sight and a massive cluster. Although some superpositions are so widely separated in redshift that one could weed them out with photometric redshifts, a large fraction are so close together that
photometric redshifts are a useless discriminant.
We degrade the SHELS redshift survey to produce 
a ``perfect'' set of 3\% photometric redshifts. We then examine weak lensing peaks ranked
3 (a superposition; Figure \ref{fig:4panel5vs3.ps} ) and 5 (a cluster at $z = 0.3$; Figure
\ref {fig:4panel5vs3.ps}) as an example of the impact of
photometric redshifts of typical accuracy on the assessment of cluster detection.

Typical photometric redshifts for luminous red galaxies are accurate to $\Delta{z}/(1 + z) \sim
0.03$ (Padmanabhan et al. 2005; Kurtz et al. 2007).  To simulate the result of a photometric redshift survey of F2 to the same depth as SHELS we simply replace the redshift for each galaxy with
a photometric redshift drawn from a Gaussian distribution with a dispersion
of  $0.03(1+z)$. Obviously these photometric redshifts are perfect;
we do not simulate the impact of systematic error and we ignore the difficulty of obtaining accurate photometric redshifts for bluer objects. 

As a telling example of the impact of photometric redshifts, we revisit weak lensing peaks 3 and 5 (Figure \ref{fig:4panel5vs3.ps}).  Figure \ref{fig:fake3and5.ps} again shows the SHELS redshift distribution for peaks 3 and 5, but here we focus on the redshift range 0.2 to 0.4 (panels labeled Peak 3 and Peak 5). The bins are 250 km s$^{-1}$.
The difference between the superposition
along the peak 3 line-of-sight and the cluster along the peak 5 line-of-sight is obvious.  Companion panels show the simulated 3\% photometric redshift distribution along the lines-of-sight toward peaks 3 and 5. The two redshift distributions with perfect photometric redshifts are nearly indistinguishable. There are 111 galaxies with R$<$20.6 in
the redshift interval for Peak 3; there are 86 for Peak 5.  
Thus the 
distribution of simulated photometric redshifts for peak 3 appears somewhat more impressive even though the
galaxies lie in concentrations with smaller velocity dispersion (150 - 340 km s$^{-1}$) than the cluster along the line-of-sight toward peak 5 (729 km s$^{-1}$).

Photometric redshifts dilute real clusters and make superpositions look like clusters
if the redshift range covered by the superposition is comparable with or less than the accuracy of the photometric redshifts. Because the uncertainty in a photometric redshift is large compared to the velocity dispersion of a cluster (or to the scale of voids in a redshift survey), the
resulting distribution of photometric redshifts indicates the number of galaxies in a redshift range rather than  the more physical quantity, the velocity dispersion of 
a system or systems of galaxies. 
We conclude that the use of photometric redshifts compromises the distinction between clusters and
superpositions of several lesser systems along the line-of-sight. Thus, without spectroscopic redshifts, an
excessive number of weak lensing peaks appear to correspond to ``clusters''.

\section {Discussion}

\label{discussion}

Combining the SHELS redshift survey with the DLS weak lensing map provides an assessment of both the efficiency and completeness of weak lensing in identifying massive clusters. As demonstrated by the theoretical models of Hamana et al. (2004), the efficiency and completeness of a catalog of massive clusters derived from weak lensing are a strong function of the cutoff S/N used to identify peaks in the weak lensing map. Maturi et al. (2009) add further weight to this conclusion. Hamana et al. (2004) and Maturi et al. (2009) conclude that minimum S/N cutoffs in the range 3-5 give an optimal balance between completeness and efficiency. Both Hamana et al (2004) and Hennawi \& Spergel (2005) show that there are intrinsic limitations, independent of
observational issues, in both the efficiency of weak lensing cluster identification and in the completeness of cluster catalogs derived from weak lensing.

Dietrich et al. (2008) also use ray-tracing simulations to gain understanding of the efficacy of convergence maps for massive halo detection. They calculate  the cumulative distribution of offsets between the positions of
peaks in the convergence map and the centers of massive halos. They show that 75\% of the offsets are 
less than 2.15$^{\prime}$. This offset is larger than the typical offset we find between peaks in the
DLS map and SHELS systems: we find a mean offset of 0.5$^{\prime}$ and a maximum of 1.5$^{\prime}$. The redshift range of the SHELS clusters coincident with DLS peaks is similar to the range considered by
Dietrich et al. (2008), but different weighting with redshift may contribute to the difference in offsets. 

Comparison of the catalog of systems with $\sigma \gtrsim 500$ km s$^{-1}$ derived from SHELS with the
significant peaks in the DLS map provides estimates of {\it both} the efficiency and completeness of the
set of systems detected in the weak lensing map. To our knowledge this comparison extending to low significance weak lensing peaks is the first of its kind. We can compare the measures of efficiency and completeness with the ray-tracing simulations of Hamana et al. (2004). We can also compare the completeness
of weak lensing detections as a function of redshift.
Next, we briefly review the conclusions of Hamana et al. (2004) and  compare them with our results.

Hamana et al. (2004) used analytic calculations and ray-tracing through large n-body simulations to explore the effectiveness of weak lensing surveys in identifying massive clusters of galaxies. They show that 
$\nu \gtrsim 4$ gives an optimal balance between efficiency and completeness in cluster identification.
Their fiducial survey has $\sim 30$ sources arcmin$^{-2}$ with the typical source as $z \sim$ 1 in contrast with the DLS source density of 23 sources arcmin$^{-2}$ and typical source redshift of 
0.7 --- 0.8. In simulated weak lensing maps with noise properties similar to the observations, Hamana et al. (2004) calculate that the completeness is $\sim$63\%
and the efficiency is $\sim$ 37\% for $\nu > 4$. These statistics apply only to clusters which should be detected at or above the weak lensing threshold. They omit clusters which should appear below the threshold but are boosted by noise. This boosting increases the apparent efficiency by $\sim 20$\%. Hamana et al. (2004) emphasize the fascinating and surprising result that noisy maps boost the completeness of cluster detection at the expense of efficiency (Figure 16 of their paper).

We can use Figures \ref{fig:ranksigma.ps} and \ref{fig:sigmavsz.ps}  to make a rough comparison of our results with the Hamana et al. (2004)
simulations. If we cut both the SHELS cluster sample (open symbols in Figure \ref{fig:sigmavsz.ps}) and the DLS detections (solid symbols in Figure \ref{fig:sigmavsz.ps})
at $\nu > 4$ (upper dotted sensitivity curve), the completeness is $\sim 60$\% and the efficiency
(from Figure \ref {fig:ranksigma.ps}) is $\sim 43$\%. Here we include Peak 5 (the predicted lensing signal is well within the uncertainty of the
$\nu =4$ threshold) as a DLS/SHELS detection; we do not include 
x-ray cluster (solid star) in our evaluations of efficiency/completeness because detection requires a higher resolution map.  Although there are only 7
weak lensing peaks with $\nu > 4$ and 5 clusters from SHELS which contribute to this assessment, we see the
trend emphasized by Hamana et al. (2004); the fractional completeness of cluster identification exceeds the efficiency. If we make the comparison at $\nu = 3.5$ we have 12 weak lensing peaks and 8 SHELS clusters (again ignoring the x-ray cluster indicated by the solid star in Figure \ref{fig:sigmavsz.ps}). The completeness is $\sim 50$\% and the efficiency
is $\sim 33$ \%. Within the considerable uncertainties of the comparison, these results are  consistent with the predictions of Hamana et al. (2004).

The DLS efficiency for cluster identification is consistent with the results of Schirmer et al. (2007) but inconsistent with the much greater efficiencies claimed by Miyazaki et al. (2007), Gavazzi \& Soucail (2007) and Kubo et al. (2009). A major reason for a lower efficiency is rejection of systems with a velocity dispersion too low to account for the weak lensing signal; this rejection is obviously impossible without a dense redshift survey like SHELS. Another contributing factor in the comparison with Kubo et al. (2009) is the smaller offset we allow between the position of the lensing peak and the SHELS system.  

We can push the comparison between the theory and our survey a bit further by considering detections 
as a function of redshift. In making this comparison, we take the greater mass threshold at greater redshift
(see Figure \ref{fig:sigmavsz.ps}) into
account. Following the same approach, Hamana et al. (2004) find that there is no strong dependence of completeness on redshift (their Figure 20).
In our survey, there is a one-to-one correspondence between the three lensing peaks and SHELS clusters at the peak sensitivity of both surveys ($z \sim 0.3$). At greater redshift the weak lensing survey is woefully incomplete although two of the five clusters which should be detected at $\nu > 3.5$ correspond to peaks in the weak lensing map with $\nu > 1.65$. The reasons for the non-detections at greater redshift are unclear, but may include the structure of the 
clusters, the details of the large-scale structure along the line-of-sight, and the properties of the noise in the DLS map.  Small number statistics are also an obvious issue here, but the comparison between theory
and observation certainly indicates
issues worth further exploration.

\section {Conclusion}

We use a four square degree region of the DLS to compare the detection of massive clusters with a weak lensing convergence map and with a dense foreground redshift survey, SHELS. This comparison, based on a
dense redshift survey completely covering the weak lensing field is the first of its kind.

We calculate the sensitivity of the DLS as a function of redshift to clusters of a particular rest frame line-of-sight velocity dispersion. We use this sensitivity curve to evaluate both the efficiency and completeness of the set of ``projected mass selected'' clusters corresponding to convergence peaks with signal-to-noise, $\nu \geq 3.5$. We conclude that the efficiency is 33\% and the completeness is 50\% for 
clusters more massive than $\sim 1.7\times 10^{14}$M$_\odot$ (rest frame velocity dispersion $\gtrsim 600$ km s$^{-1}$). These results agree with the more pessimistic previous evaluation of the efficacy of
weak lensing by  Schirmer et al. (2007) and they are consistent with ray tracing simulations by Hamana et al. (2004).  

We examine the coincidence between DLS convergence map peaks with $\nu \gtrsim 1$ and clusters in SHELS with rest frame line-of-sight velocity dispersion $\gtrsim$ 500 km s$^{-1}$. For all nine coincidences, the offset between the position of the weak lensing peak and the cluster center in SHELS is less than 1.5$^\prime$. These small offsets are consistent with previous observations by Gavazzi \& Soucail (2007), but smaller than the theoretical predictions of Dietrich et al. (2008). We use these offsets and the complete SHELS catalog to demonstrate that these coincidences are probably real detections even at low signal-to-noise; we note that two of these coincidences are SHELS clusters that should be detected in the DLS at greater significance. These latter coincidences indicate that 
failure to detect SHELS clusters in the DLS may be related to the properties of the noise in the $\kappa$ map.
We show that, as expected, the efficiency of the weak lensing survey for cluster detection decreases steeply with signal-to-noise.

X-ray observations cover only a small portion of the DLS field, thus limiting the evaluation of the efficiency and completeness of the DLS relative to an x-ray selected catalog. There are seven known extended cluster x-ray sources in the DLS field. Among these, three are DLS detections. The SHELS velocity dispersion indicates that the x-ray cluster XMMU J091935+303155 should be detected in the DLS, but it is not. SHELS provides velocity dispersions for the remaining three systems and suggests that none of them should produce a
$\nu \geq 3.5$ peaks in the weak lensing map without significant boosting from large-scale structure superposed along the line-of -sight.

The SHELS/DLS weak lensing efficiency for cluster selection is more pessimistic than previous surveys at least in part because we require that the velocity dispersion measured in SHELS be adequate to produce the lensing signal. Previous evaluations of efficiency based on counts and/or photometric redshifts by necessity include superpositions of systems with low velocity dispersion. We use SHELS to construct an example 
demonstrating the failure of photometric redshifts to distinguish between a massive cluster and a superposition of groups along the line-of-sight.

The underlying causes of the low efficiency and completeness of the weak lensing candidate list are
hard to identify in our sample although the properties of the noise in the $\kappa$ map are probably important.
Although the clusters the DLS detects in SHELS  are richer than the ones the DLS does not detect at redshifts of 0.4 ---0.55, this difference in optical properties is unlikely to account for the non-detections. There is also no obvious reason for the DLS failure to detect the x-ray cluster XMMU J091935+303155. 
Although the the conclusions of the DLS/SHELS comparison are obviously limited by small number statistics, they demonstrate that more extensive evaluation of both the efficiency and completeness of weak lensing 
cluster candidate lists relative to other catalogs is crucial for optimal application of 
weak lensing selected cluster catalogs to cosmology.

\acknowledgments

We thank P. Berlind and M. Calkins for their expert operation
of the Hectospec.  D. Mink, J. Roll, S. Tokarz, and W. Wyatt
constructed and ran the Hectospec pipeline. We thank H. Khiabanian and J. Kubo for 
guidance in using the DLS convergence maps. We thank the anonymous referee for
insightful, careful, gracious comments
that inspired us to improve the paper.  We also thank S.
Kenyon and M. Zaldarriaga for many insightful discussions. The Smithsonian
Institution generously supported Hectospec and SHELS.
Lucent Technologies and NSF grants AST 04-41-72 and AST
01-34753 generously supported the DLS. We appreciate generous allocations
of telescope time on the Kitt Peak National Observatory
4 m telescope and on the MMT. I. Dell'Antonio is supported by NSF-AST grant AST-0708433.

{\it Facilities:}\facility {MMT(Hectospec)},\facility{ Mayall(MOSAIC-I and II widefield cameras)}

%

\clearpage
\begin{landscape}
\begin{deluxetable}{lccccccccccc}
\tablecolumns{12}
\tablewidth{0pc}
\tabletypesize{\footnotesize}
\tablenum{1}
\tablecaption {Velocity Dispersions of SHELS/DLS Candidate Clusters}
\tablehead{
\colhead {SHELS Name}&
\colhead {DLS Rank}&
\colhead {$\nu_{DLS}$}&
\colhead {RA$_{2000}$}&
\colhead {DEC$_{2000}$}&
\colhead {z}&
\colhead {$\sigma_{los,3}$ }&
\colhead {N$_3$}&
\colhead {err$_3$}&
\colhead {$\sigma_{los,6}$}&
\colhead {N$_6$}&
\colhead {err$_6$} \\  & & & & & & 
\colhead {km/s}& &
\colhead {km/s}&
\colhead {km/s}& &
\colhead {km/s}

}
\startdata

SHELS J0915.1+2954&-     & -     & 9:15:03.5 & 29:54:09 & 0.1319 & 565 & 11 & 134 & 505 & 16 & 130 \\
SHELS J0916.0+3028&10    & 3.76  & 9:15:57.1 & 29:49:42 & 0.1844 & 500 & 17 & 99 & 565 & 38 & 71 \\
SHELS J0916.2+2949$^D$&10    & 3.76  & 9:16:10.9 & 29:48:44 & 0.5343 & 880 & 11 & 127 & 867 & 29 & 71 \\
SHELS J0916.3+2916&17    & 2.69  & 9:16:19.2 & 29:15:47 & 0.5347 & 587 & 8 & 74 & 642 & 14 & 70 \\
SHELS J0916.7+2920&34    & 1.97  & 9:16:40.1 & 29:19:52 & 0.2158 & 595 & 16 & 75 & 569 & 26 & 75 \\
SHELS J0916.8+2908&-     & -     & 9:16:50.0 & 29:08:19 & 0.3356 & 591 & 7 & 61 & 532 & 10 & 61 \\
SHELS J0916.9+3003&-     & -     & 9:16:56.7 & 30:03:08 & 0.3189 & 487 & 18 & 68 & 442 & 42 & 49 \\
SHELS J0918.1+3038$^D$&42    & 1.65  & 9:18:05.8 & 30:37:48 & 0.3970 & 686 & 10 & 147 & 749 & 21 & 101 \\
SHELS J0918.2+3057$^D$&-     & -     & 9:18:09.8 & 30:56:56 & 0.4244 & 667 & 9 & 125 & 732 & 16 & 124 \\
SHELS J0918.3+3024&-     & -     & 9:18:16.0 & 30:24:07 & 0.1241 & 539 & 25 & 74 & 534 & 53 & 53 \\
SHELS J0918.6+2953$^D$&5     & 4.48  & 9:18:38.6 & 29:53:22 & 0.3178 & 729 & 41 & 90 & 676 & 66 & 63 \\
SHELS J0919.6+3032$^{*D}$&-     & -     & 9:19:33.3 & 30:31:59 & 0.4273 & 596 & 11 & 107 & 718 & 19 & 113 \\
SHELS J0920.1+3010&-     & -     & 9:20:03.6 & 30:10:06 & 0.4263 & 507 & 8 & 81 & 541 & 12 & 68 \\
SHELS J0920.4+3030$^{*D}$&2     & 5.72  & 9:20:22.5 & 30:30:29 & 0.3004 & - & - & - & 929 & 219 & 200 \\
SHELS J0920.9+3029$^{*D}$&1     & 6.64  & 9:20:55.6 & 30:28:38 & 0.2915 & - & - & - & 856 & 132 & 200 \\
SHELS J0921.0+2942&-     & -     & 9:20:59.6 & 29:42:00 & 0.2964 & 463 & 28 & 75 & 551 & 49 & 50 \\
SHELS J0921.2+3028$^{*}$&-     & -     & 9:21:12.7 & 30:28:08 & 0.4265 & 754 & 22 & 92& 772 & 40 & 88 \\
SHELS J0921.3+2946&-     & -     & 9:21:13.9 & 29:45:37 & 0.3834 & 662 & 11 & 150 & 549 & 18 & 131 \\
SHELS J0921.4+2958$^D$&15    & 2.87  & 9:21:24.9 & 29:58:12 & 0.4318 & 597 & 15 & 94 & 818 & 24 & 109 \\
SHELS J0923.6+2929&33    & 1.97  & 9:23:38.0 & 29:28:35 & 0.2216 & 493 & 23 & 68 & 454 & 40 & 46 \\
\enddata
\\
\tablenotetext{*}{These clusters are the known extended x-ray sources XMMU J091935+303155, CXOU J092026+302938, CXOU J092053+302800, and CXOU J092110+302751.They also appear in Table \ref{tbl:Xrays}}
\tablenotetext{D}{The SHELS velocity dispersion implies that these clusters should be detected at $\nu_{DLS}\geq 3.5$ with the
resolution of the DLS map we use. 
See Figure \ref{fig:sigmavsz.ps}}
\label{tbl:VDisp1}
\end{deluxetable}
\end{landscape}

\clearpage
\begin{landscape}
\begin{deluxetable}{lcccccccc}
\tablecolumns{9}
\tablewidth{0pc}
\tabletypesize{\footnotesize}
\tablenum{2}
\tablecaption {Properties of Known SHELS/DLS X-Ray Clusters}
\tablehead{
\colhead {X-ray Name}&
\colhead {DLS Rank}&
\colhead {RA$_{2000}$}&
\colhead {DEC$_{2000}$}&
\colhead {z}&
\colhead {$\sigma_{los,3}$ }&
\colhead {N$_3$}&
\colhead {err$_3$}&
\colhead {Xray Flux$^a$}\\  
& & & & & 
\colhead {km/s}& &
\colhead {km/s}&
\colhead {10$^{-14}$erg\ s$^{-1}$cm$^{-2}$}

}
\startdata 

CXOU J091551+293637 & 11& 9:15:51.8 & 29:36:37 & 0.5312 & 361 & 11 & 70 &    1.80 \\
--- & 11$^b$ & 9:15:51.8 & 29:36:37 & 0.1851 & 420 & 8 & 144 &    1.80 \\
CXOU J091554+293316 & 8$^b$ & 9:15:54.4 & 29:33:16 & 0.1847 & 526 & 8 & 105 &    0.71 \\
CXOU J091601+292750 & - & 9:16:01.1 & 29:27:50 & 0.5319 & 523 & 7 & 126 &    1.80 \\
--- & -$^b$ & 9:16:01.1 & 29:27:50 & 0.1834 & 593 & 5 & 71 &    1.80 \\
XMMU J091935+303155 & - & 9:19:35.0 & 30:31:55 & 0.4273 & 596 & 11 & 107 &    - \\
CXOU J092026+302938 & 2 & 9:20:26.4 & 30:29:39 & 0.3004 & ~929$^c$ & ~219$^c$ & ~200$^c$ &   64.20 \\
CXOU J092053+302800 & 1 & 9:20:53.0 & 30:28:00 & 0.2915 & ~856$^c$ & ~132$^c$ & ~200$^c$ &   11.60 \\
CXOU J092110+302751 & - & 9:21:10.3 & 30:27:52 & 0.4265 & 754 & 222 & 92 &    9.47 \\

\enddata
\tablenotetext{a}{{\it Chandra} 0.5-2 keV flux from Table 3 of Wittman et al. (2006)}
\tablenotetext{b} {For these entries we list the number of galaxies within the intercepted large-scale structure. The velocity dispersion refers to the effective velocity width of these structures rather than to a centrally concentrated group or cluster.}
\tablenotetext{c}{The line-of-sight velocity dispersions and errors derived from the 6$^\prime$
samples as in
Table \ref{tbl:VDisp1}}
\label{tbl:Xrays}
\end{deluxetable}
\end{landscape}
\clearpage

\begin{figure}[htb]
\centerline{\includegraphics[width=7.0in]{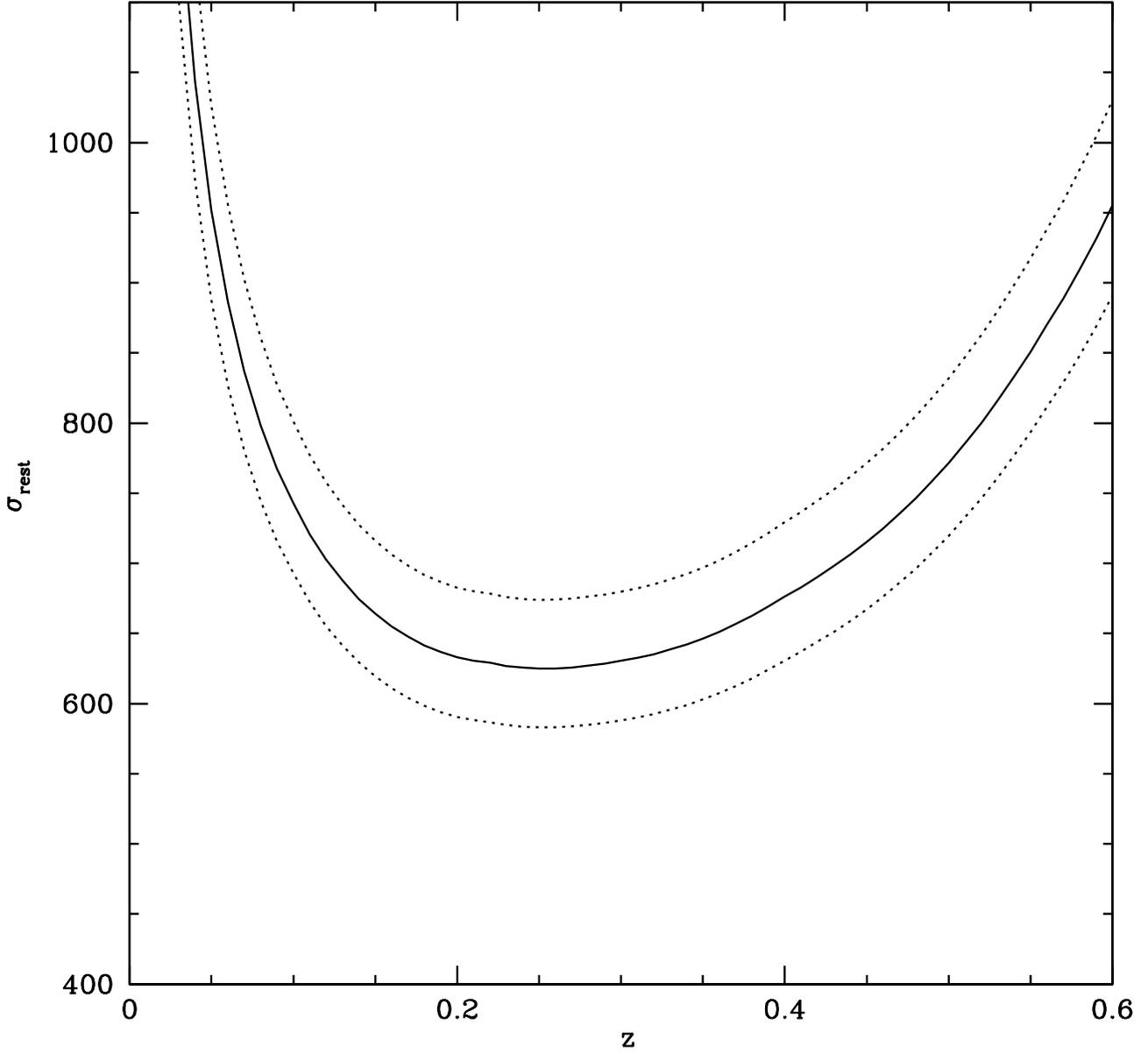}}
\vskip -5ex
\caption{DLS sensitivity as a function of redshift. The solid curve shows
the $\nu = 3.5$ detection limit. The dashed curves $\nu = 3$ (lower curve)
and $\nu = 4$ (upper curve).
\label{fig:sensitivity_band.ps}}

\end{figure}

\begin{figure}[htb]
\centerline{\includegraphics[width=7.0in]{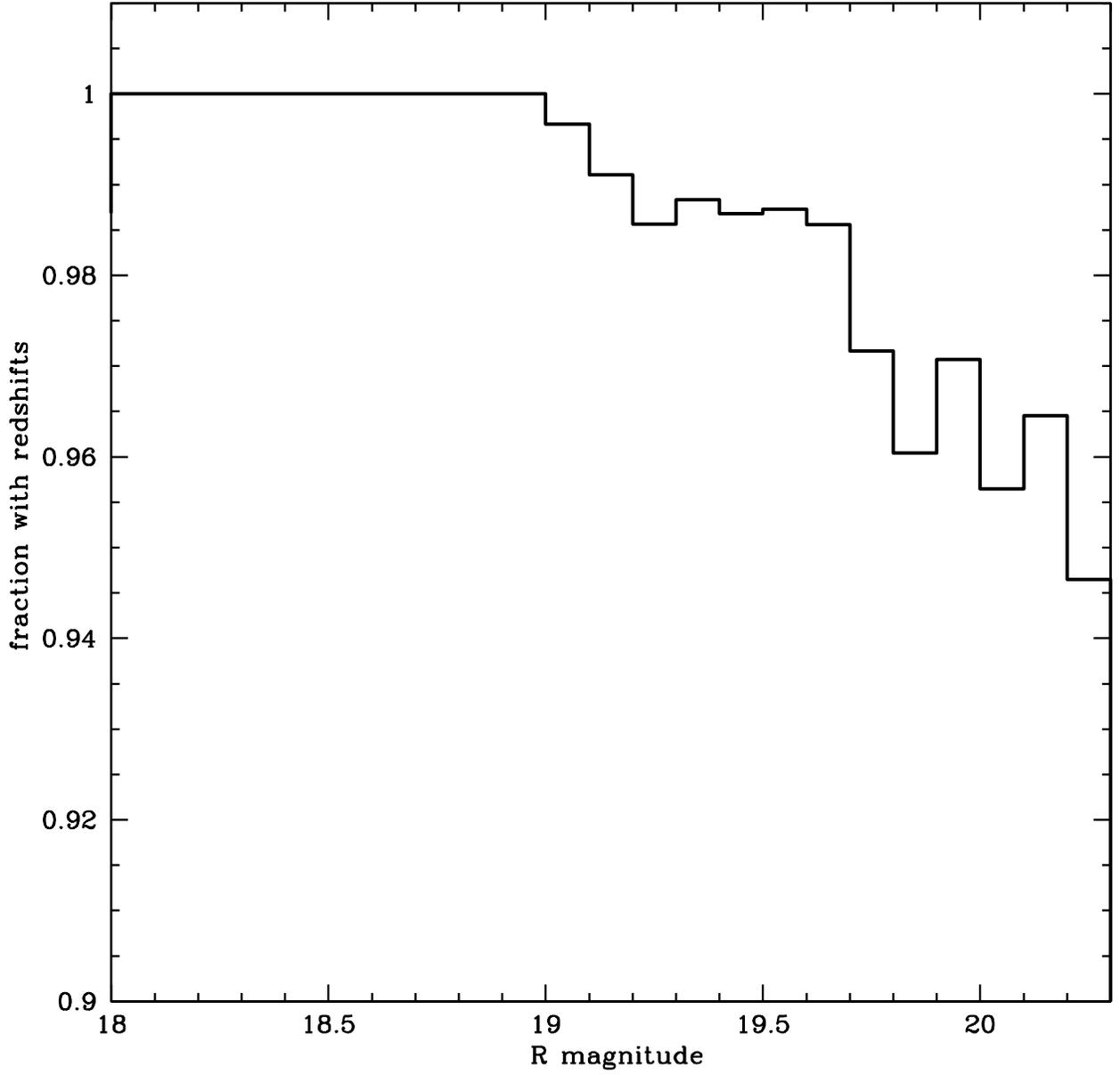}}
\vskip -5ex
\caption{Completeness of the SHELS redshift survey. The redshift survey contains
9825 galaxies with R $\leq$ 20.3. The integral completeness to R = 20.3
is 97.8\%. 
\label{fig:completeness}}
\end{figure}

\begin{figure}[htb]
\centerline{\includegraphics[width=7.0in]{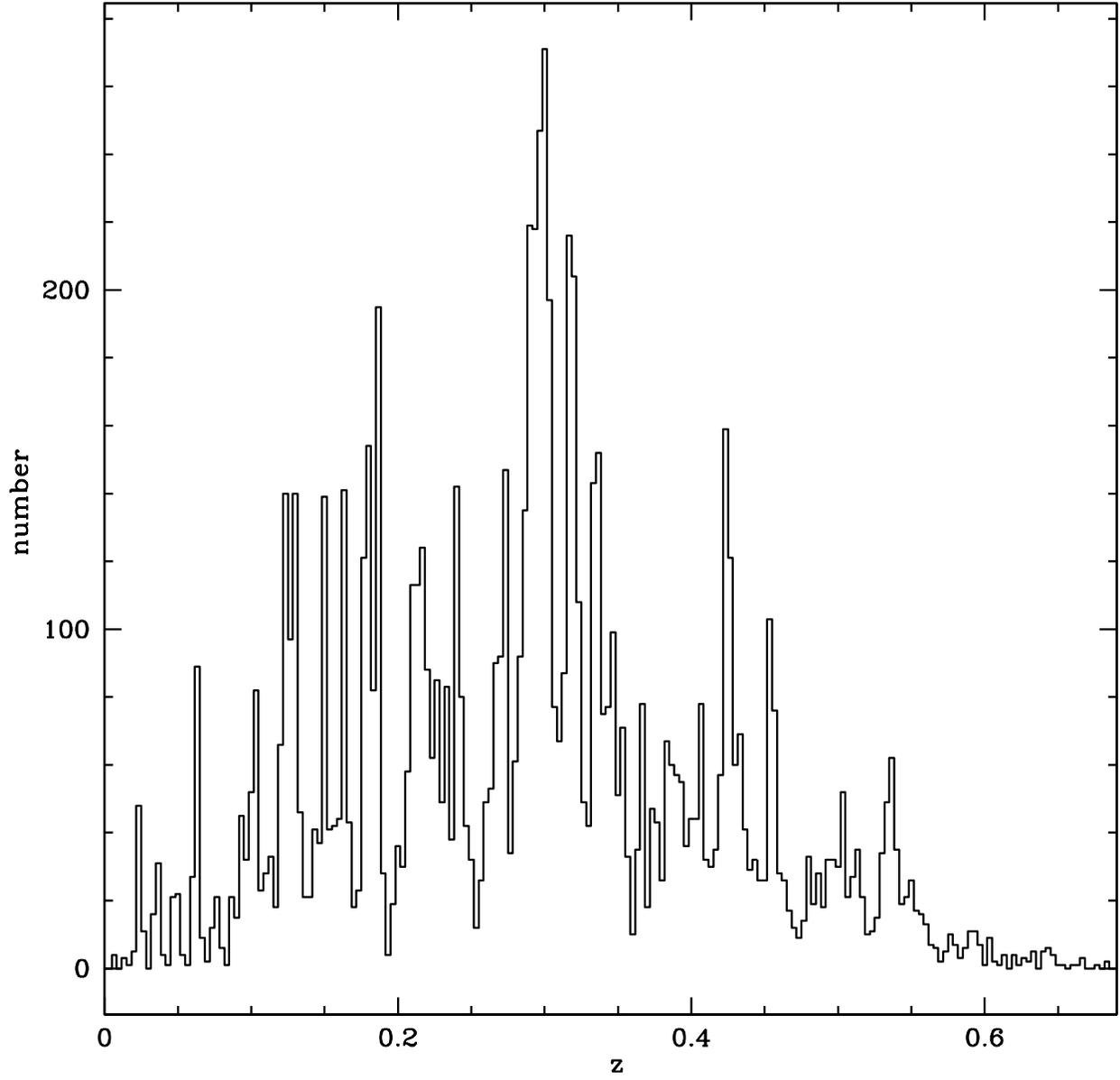}}
\vskip -5ex
\caption{Redshift distribution for the SHELS redshift survey. The peaks are the standard signature of large-scale structure.
\label{fig:zhisto}}
\end{figure}

\begin{figure}[htb]
\centerline{\includegraphics[width=7.0in]{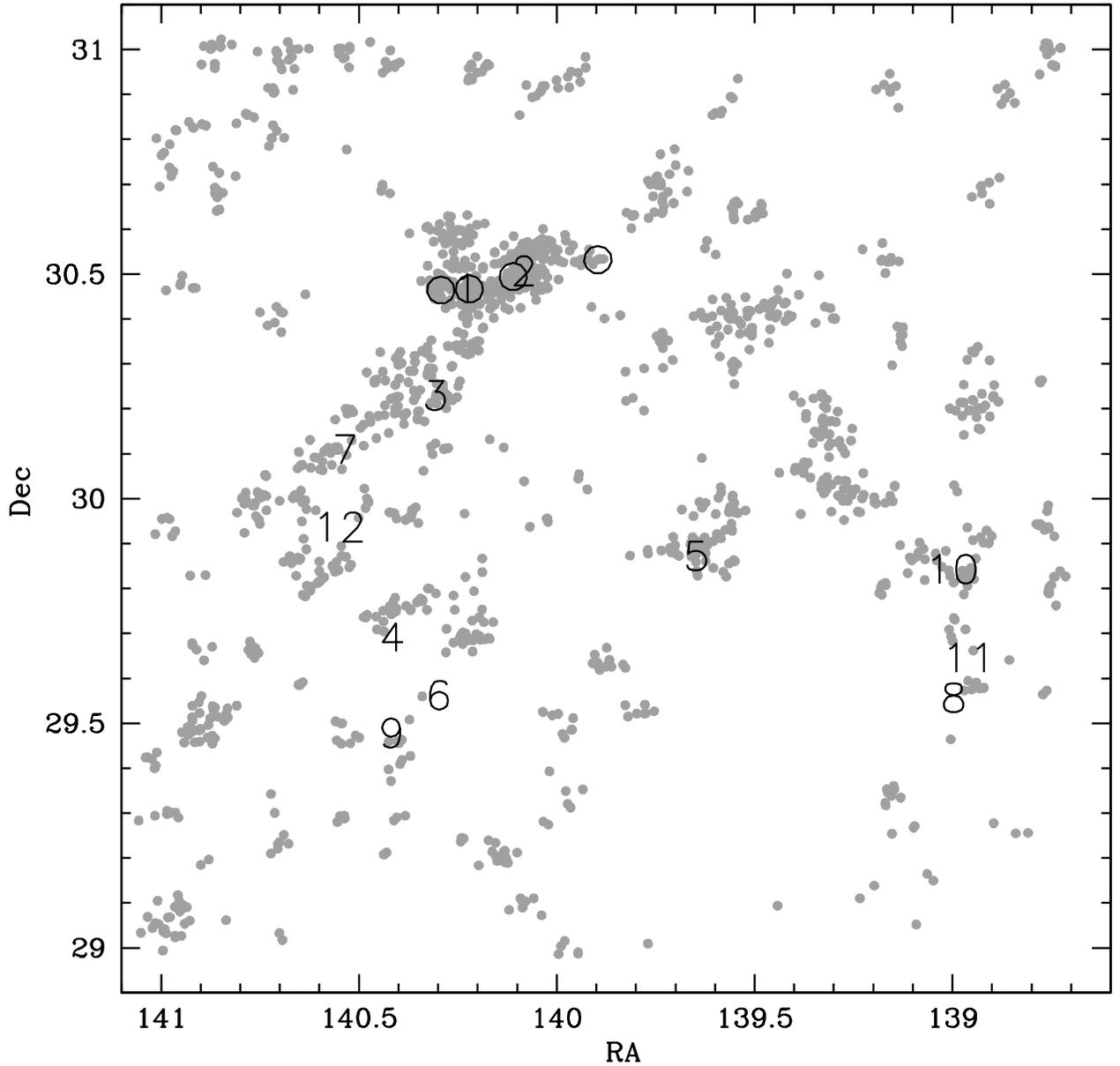}}
\vskip -5ex
\caption{Map of F2 centers of the 3$^{\prime}$ radius 5$\sigma_{SH}$ probes (gray dots) showing, x-ray
clusters at $z = 0.28-0.43$ (open circles), and centers of 
significant weak lensing peaks (numbers). The value of the number gives the rank of the convergence map peak.
Note the general correspondence between weak lensing peaks and 5$\sigma_{SH}$ probes. Note that the tick mrks on the declination axis are 6$^\prime$ apart, the diameter of the probes. 
\label{fig:5sigmamap.ps}}
\end{figure}

\begin{figure}[htb]
\centerline{\includegraphics[width=7.0in]{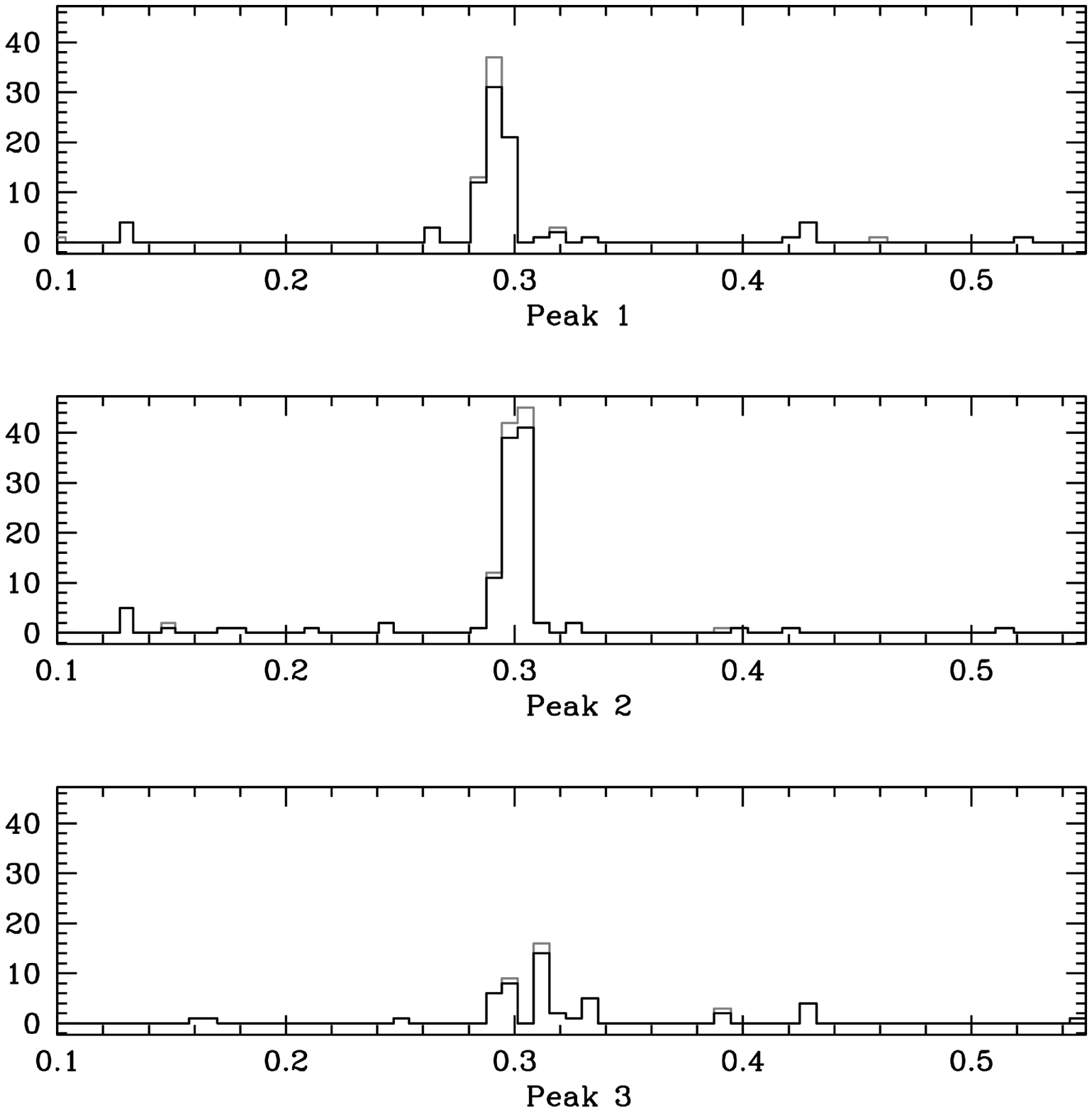}}
\vskip -5ex
\caption{Redshift histograms of galaxies  within 3$^{\prime}$ radius cones centered on
DLS weak lensing peaks ranked 1 -- 3. Bins in redshift are 0.0053(1+z).
Black histograms include galaxies with R $\leq 20.3$; gray 
histograms include galaxies with R $\leq 20.6$.
\label{fig:PeakHisto.1-3.ps}}
\end{figure}

\begin{figure}[htb]
\centerline{\includegraphics[width=7.0in]{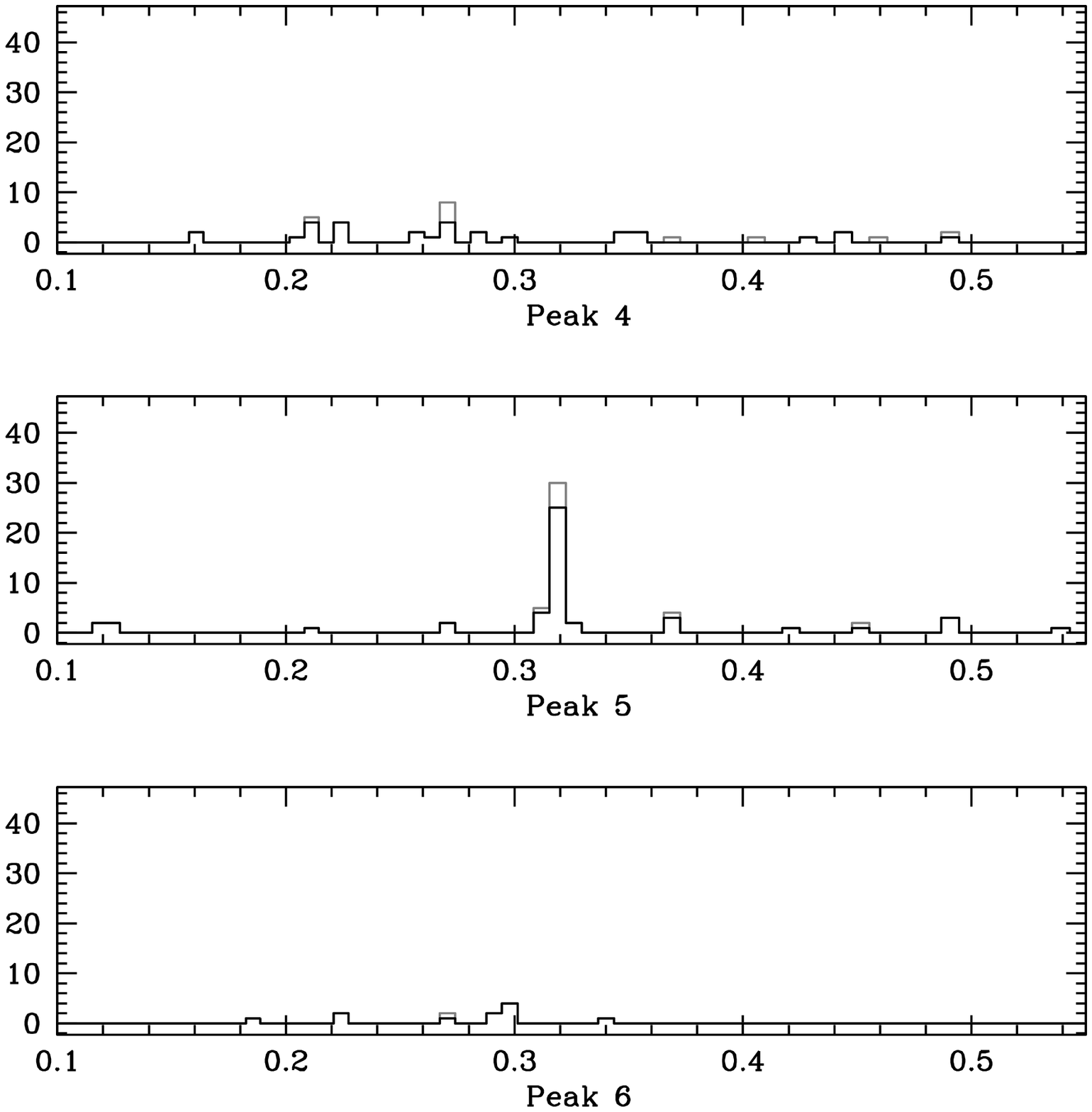}}
\vskip -5ex
\caption{Redshift histograms of galaxies within 3$^{\prime}$ radius cones centered on
DLS weak lensing peaks ranked 4 -- 6. Bins in redshift are 0.0053(1+z).
Black histograms include galaxies with R $\leq 20.3$; gray 
histograms include galaxies with R $\leq 20.6$.
\label{fig:PeakHisto.4-6.ps}}
\end{figure}

\begin{figure}[htb]
\centerline{\includegraphics[width=7.0in]{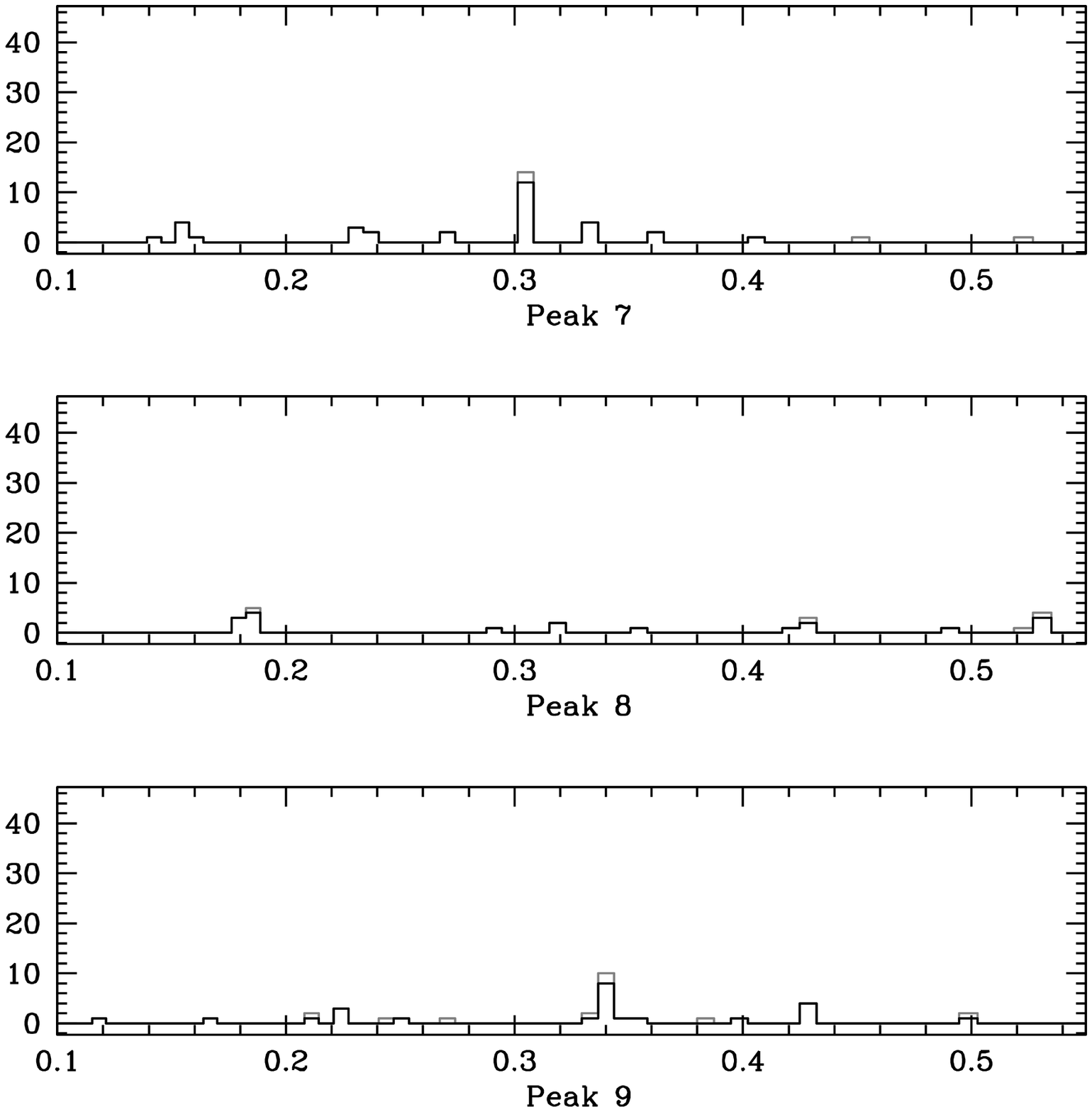}}
\vskip -5ex
\caption{Redshift histograms of galaxies within 3$^{\prime}$ radius cones centered on
DLS weak lensing peaks ranked 7 -- 9. Bins in redshift are 0.0053(1+z).
Black histograms include galaxies with R $\leq 20.3$; gray 
histograms include galaxies with R $\leq 20.6$.
\label{fig:PeakHisto.7-9.ps}}
\end{figure}

\begin{figure}[htb]
\centerline{\includegraphics[width=7.0in]{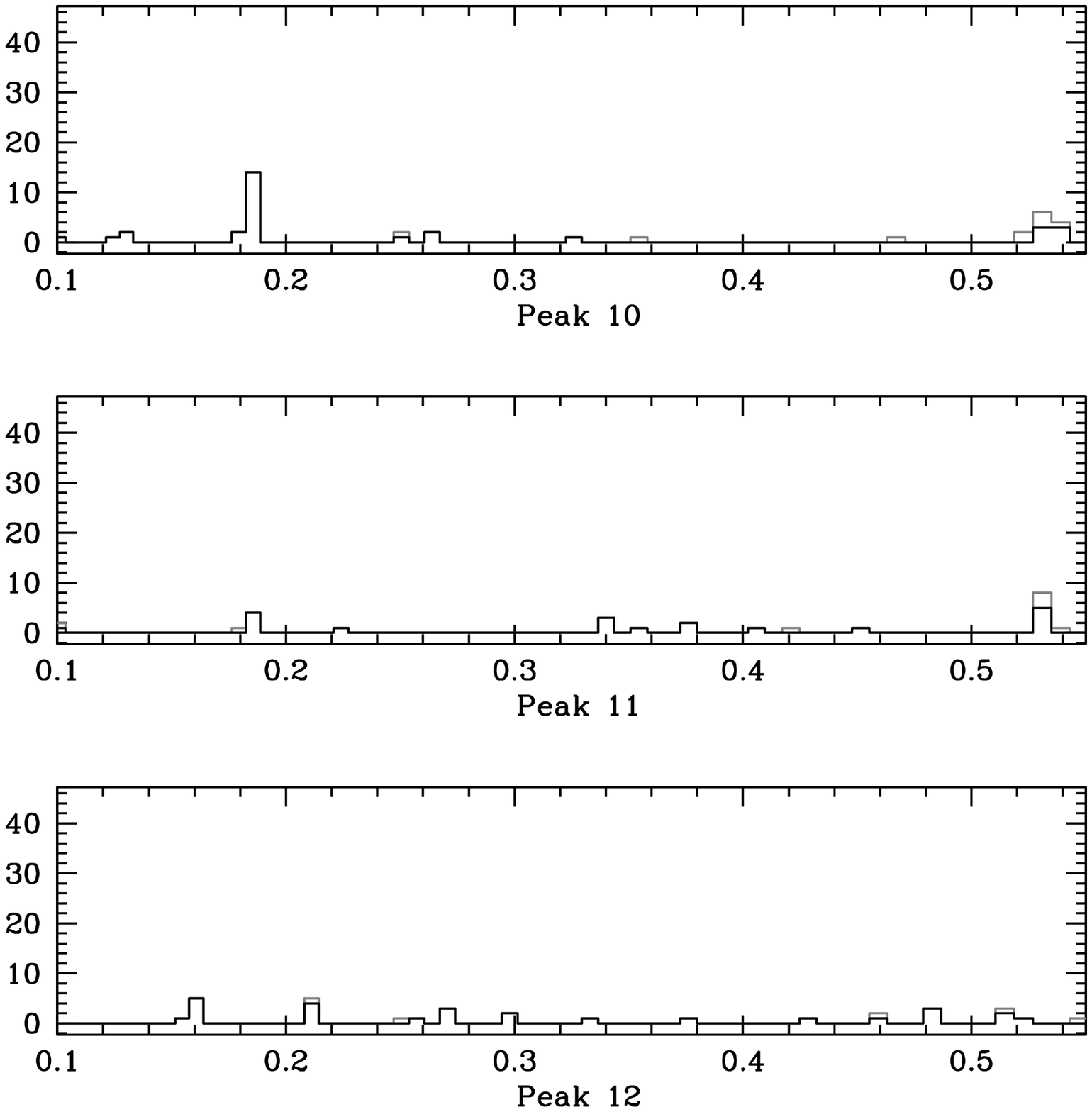}}
\vskip -5ex
\caption{Redshift histograms of galaxies within 3$^{\prime}$ radius cones centered on
DLS weak lensing peaks ranked 10 -- 12. Bins in redshift are 0.0053(1+z).
Black histograms include galaxies with R $\leq 20.3$; gray 
histograms include galaxies with R $\leq 20.6$.
\label{fig:PeakHisto.10-12.ps}}
\end{figure}

\begin{figure}[htb]
\centerline{\includegraphics[width=7.0in]{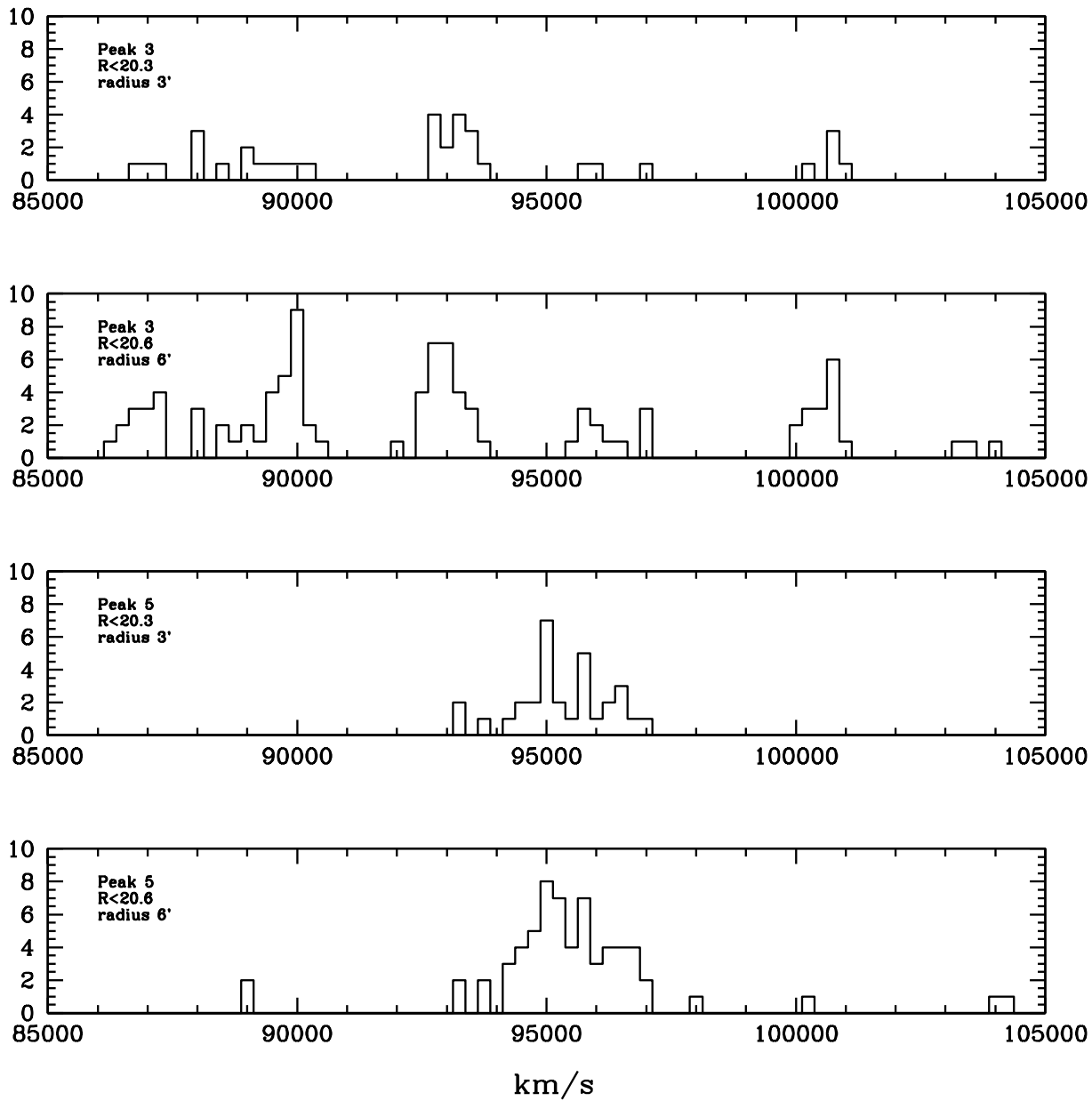}}
\vskip -5ex
\caption{Comparison of the redshift distributions of galaxies toward DLS convergence peaks ranked 3 and 5. The upper panels
show the line-of-sight redshift distribution within a 3$^\prime$ radius cone toward peak 3 for R$< 20.3$ and
within a 6$^{\prime}$ radius cone to the deeper limit R$ < 20.6$. The lower two panels show the analogous 3$^{\prime}$ 
and 6$^{\prime}$  line-of-sight redshift distributions for peak 5. Peak 5 is a rich cluster; peak 3 is a superposition.
\label{fig:4panel5vs3.ps}}
\end{figure}

\begin{figure}[htb]
\centerline{\includegraphics[width=7.0in]{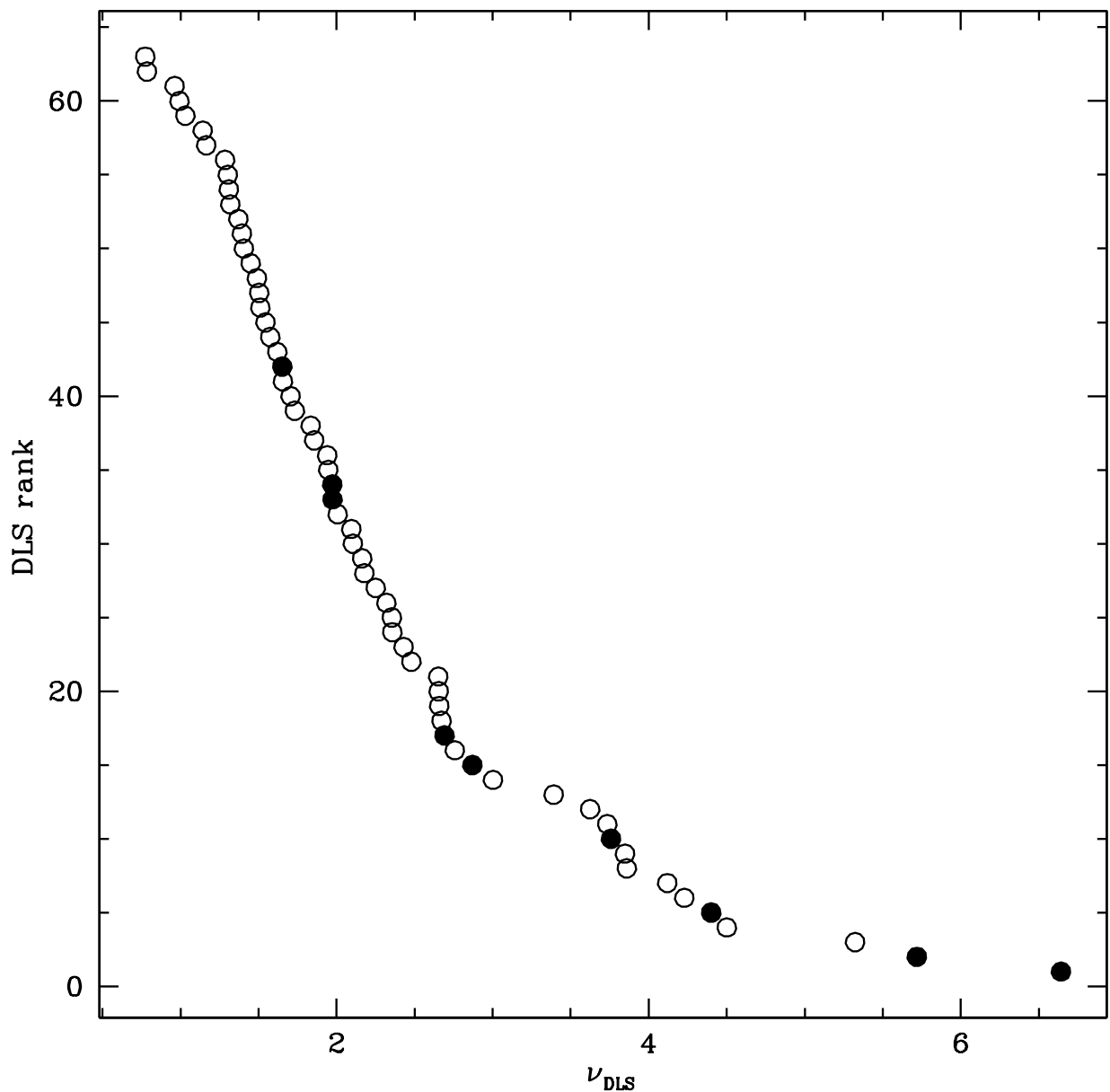}}
\vskip -5ex
\caption{Rank as a function of $\nu$ for weak lensing peaks in
the F2 field. Filled circles indicate coincidence of a SHELS cluster
with $\sigma_{los} \gtrsim 500$ km s$^{-1}$. There are no  
SHELS clusters with $\sigma_{los} \gtrsim 500$ km s$^{-1}$
along the line-of-sight toward peaks designated with empty circles. Among the 12 DLS peaks with $\nu \geq 3.5$, only 4 
correspond to massive SHELS clusters, an efficiency of 33\%.
\label{fig:ranksigma.ps}}
\end{figure}

\begin{figure}[htb]
\centerline{\includegraphics[width=7.0in]{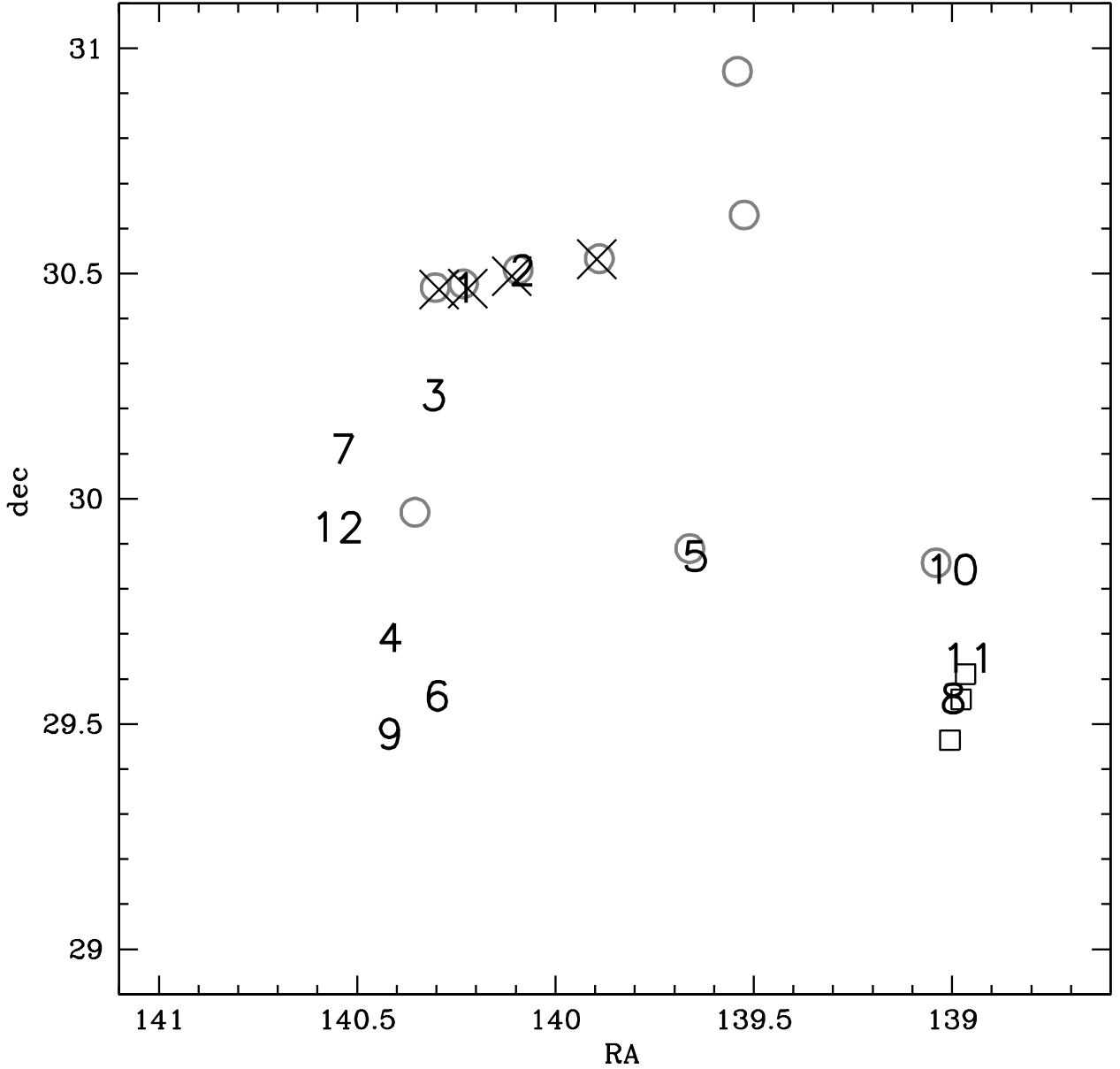}}
\vskip -5ex
\caption{Map of the DLS field showing the positions of known x-ray clusters (crosses and open boxes), significant weak lensing peaks (numbers), and candidate SHELS clusters that lie above the threshold for detection in the weak lensing map (open circles;
see Figure \ref{fig:sigmavsz.ps}). Crosses denote the clusters CXOU J092026+302938, CXOU J092053+302900, CXOU J092026+302938, CXOU J092053+302900, and XMMU J091935+303155; all of these clusters are detected in SHELS and should be detected by weak
lensing. Open boxes denote the x-ray sources CXOU J091551+293637, CXOU J091554+293316,
and CXOU J091601+292750. There is no convincing optical counterpart to the central cluster of these three; the other two
have SHELS velocity dispersions too low to place them above the lensing detection threshold in Figure \ref{fig:sigmavsz.ps}.
Open circles designate SHELS clusters ($^D$ in Table \ref{tbl:VDisp1}) that lie above the DLS detection threshold.
\label{fig:comparison_map.ps}}
\end{figure}

\begin{figure}[htb]
\centerline{\includegraphics[width=7.0in]{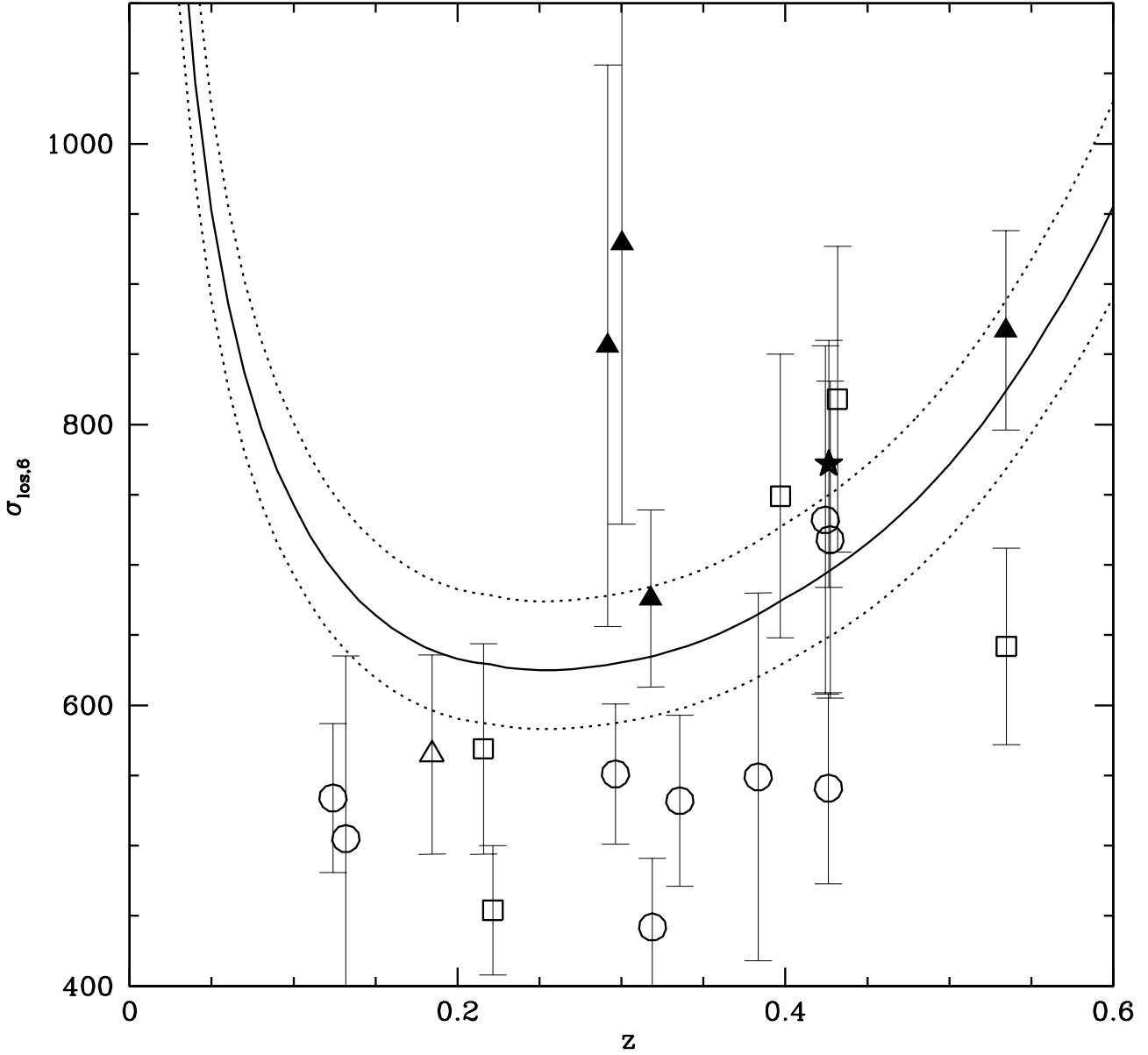}}
\vskip -5ex
\caption{SHELS clusters and DLS sensitivity. The solid curve shows the
$\nu = 3.5$ sensitivity as a function of redshift and the dotted lines indicate the $ \nu = 3$ to $\nu = 4$ range from Figure 
\ref  {fig:sensitivity_band.ps}. The 
vertical axis is rest frame line-of-sight velocity dispersion, $\sigma_{los,6}$ Filled triangles are
DLS clusters coincident with $\nu \geq 3.5$ weak lensing peaks.
Open squares are DLS systems coincident with lower DLS peaks in the range 1$\leq \nu < 3.5$. Open 
circles are SHELS clusters with no DLS counterpart. The filled star indicates 
the x-ray cluster XMMU J091935+303155 detected only in a higher resolution DLS map at $\nu = 5.7$. Among the
massive clusters that lie above the $\nu = 3.5$ threshold, the DLS detects
$\sim 50$\%. 
\label{fig:sigmavsz.ps}}
\end{figure}

\begin{figure}[htb]
\centerline{\includegraphics[height=7.5in]{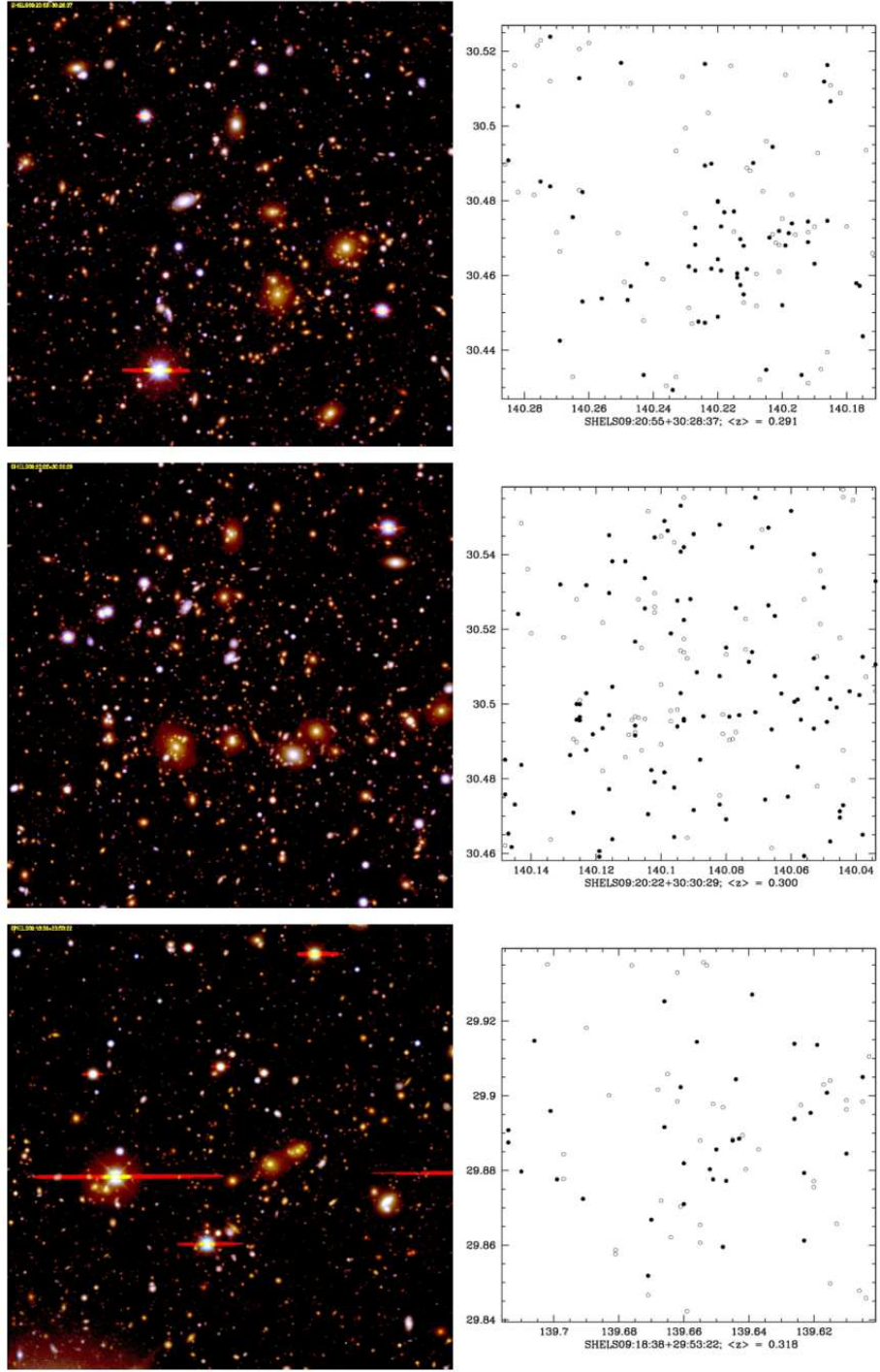}}
\vskip -2ex
\caption{DLS images of the central $6^\prime \times 6^\prime$ regions of SHELS cluster candidates that should be detected at $\nu \geq 3.5$. SHELS J0920.9+3029 (top) is
DLS peak 1; SHELS J0920.4+3030 (middle) is DLS peak 2; SHELS J0918.6+2953 (bottom) is DLS peak 5. The plots on the right correspond to the images on the left and show galaxies with redshifts in SHELS; the solid dots are system members. The redshift range for these three systems is 0.291 (top) to 0.318 (bottom).
\label{fig:Clusters1.eps}}
\end{figure}

\begin{figure}[htb]
\centerline{\includegraphics[height=7.5in]{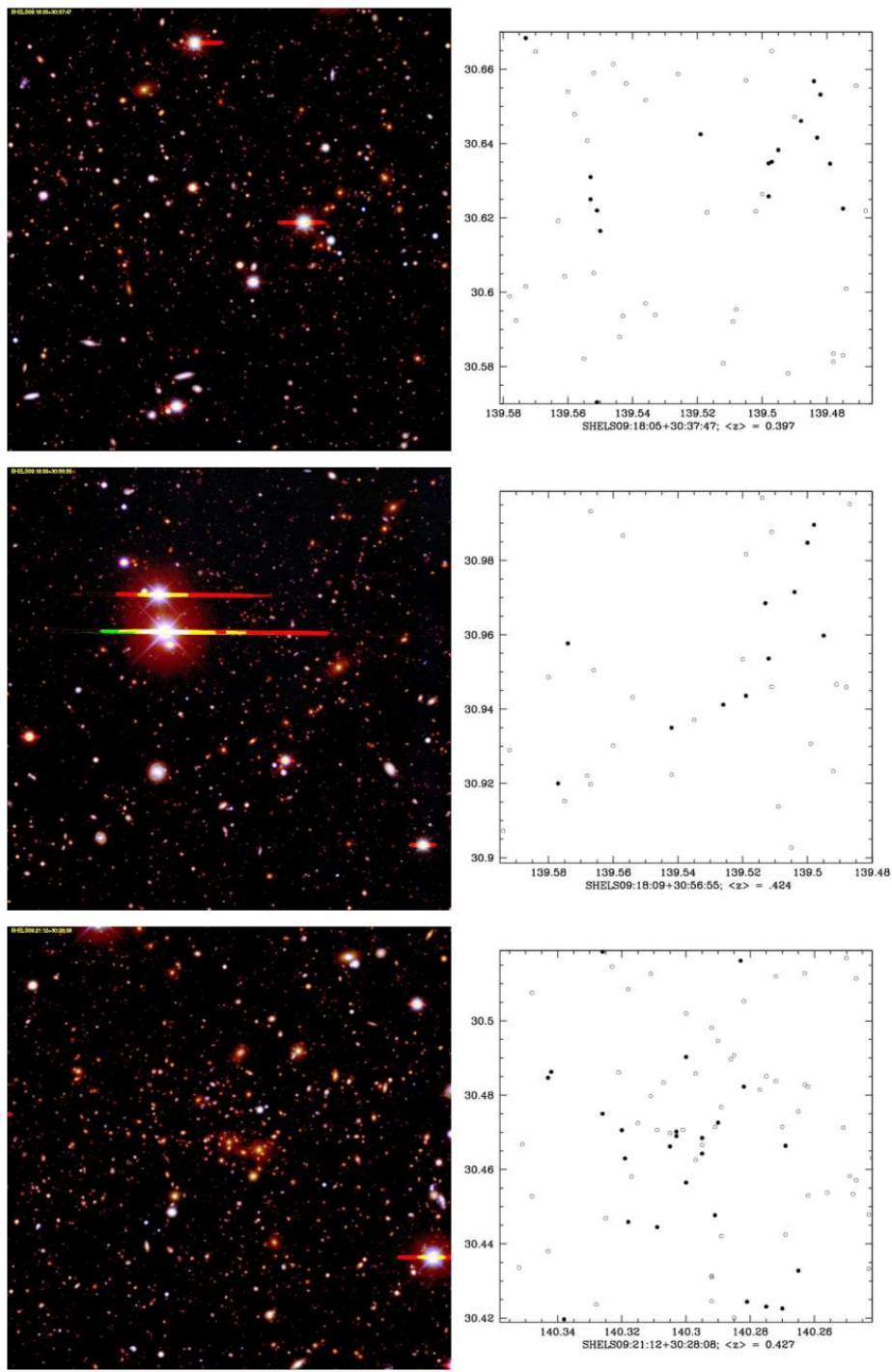}}
\vskip -2ex
\caption{DLS images of the central $6^\prime \times 6^\prime$ regions of SHELS cluster candidates that should be detected at $\nu \geq 3.5$. SHELS J0918.1+3038 (top)  is DLS peak 42; SHELS J0918.2+3057 (center) is undetected; SHELS J0921.2+3028 (bottom) is an extended x-ray source (CXOU J092110+302751) detected by the DLS in a higher resolution map. The plots on the right correspond to the images on the left and show galaxies with redshifts in SHELS; the solid dots are system members. The redshift
range for these systems is 0.397 (top) to 0.427 (bottom).
\label{fig:Clusters2.eps}}
\end{figure}

\begin{figure}[htb]
\centerline{\includegraphics[height = 7.5in]{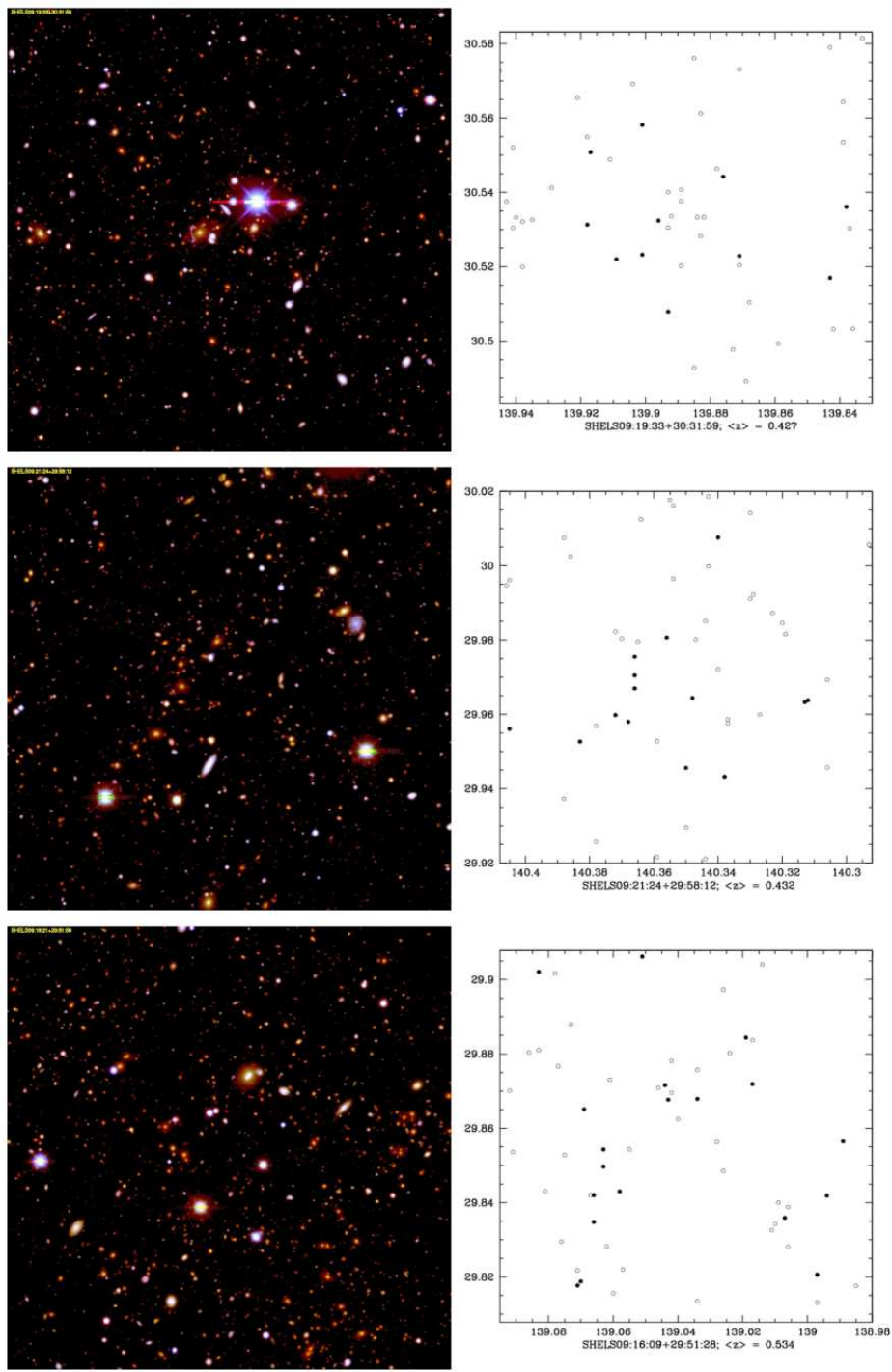}}
\vskip -2ex
\caption{DLS images of the central $6^\prime \times 6^\prime$ regions of SHELS cluster candidates that should be detected at $\nu \geq 3.5$. SHELS J0919.6+3032 (top) is an extended x-ray source (XMMU J091935+303155) but is undetected by the DLS; SHELS J0921.4+2958 (middle) is DLS peak 15; SHELS J0916.2+2949 (bottom) is DLS peak 10.  The plots on the right correspond to the images on the left and show galaxies with redshifts in SHELS; the solid dots are system members. The redshift range for these
systems is 0.427 (top) to 0.534 (bottom).
\label{fig:Clusters3.eps}}
\end{figure}

\begin{figure}[htb]
\centerline{\includegraphics[width=7.0in]{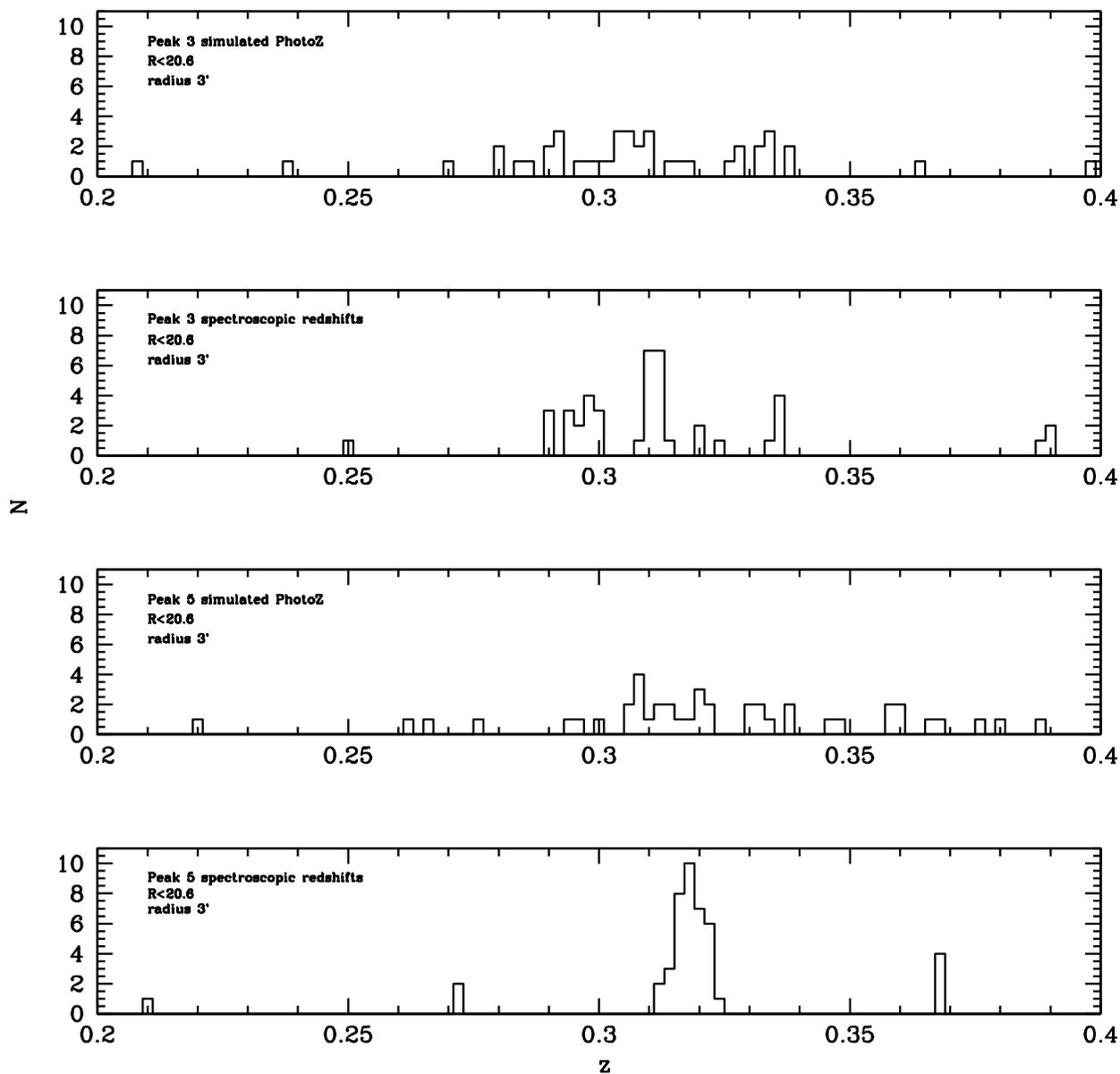}}
\vskip -5ex
\caption{Impact of photometric redshifts on the ability to discriminate between a cluster of galaxies and 
a superposition of groups along the line-of-sight. The upper two panels show the distributions of
perfect simulated photometric redshifts for convergence peaks 3 in the most populated redshift range 0.2 ---
0.4 (top) and the true redshift distribution; the lower two panels show the perfect photometric redshift distribution for the cluster along the line-of-sight toward convergence map peak 5 (upper panel) and
the true redshift distribution (lower panel). Note the similarity of the photometric redshift distributions
and the marked difference between the true redshift distributions.
\label{fig:fake3and5.ps}}
\end{figure}

\end{document}